\documentclass[aps, prc, twocolumn, showpacs, floatfix, nofootinbib, nobibnotes, superscriptaddress, amsmath, amssymb,longbibliography]{revtex4-1}

\usepackage[titletoc]{appendix}
\usepackage{mathrsfs}
\usepackage{geometry}
\usepackage{graphicx}
\usepackage{enumerate}
\usepackage{hyperref}
\usepackage{graphicx}
\usepackage{color}
\usepackage{float}
\usepackage{dcolumn}
\usepackage{multirow}
\usepackage{xfrac}
\usepackage{bm}

\newcolumntype{d}[1]{D{.}{.}{#1}}

\begin{document}

\renewcommand{\arraystretch}{1.25}

\title{Effective field theory for vibrations in odd-mass nuclei}

\author{E. A. Coello P\'erez}

\affiliation{Institut f\"ur Kernphysik, Technische Universit\"at
Darmstadt, 64289 Darmstadt, Germany}
\thanks{present address}

\affiliation{ExtreMe Matter Institute EMMI, Helmholtzzentrum f\"ur
Schwerionenforschung GmbH, 64291 Darmstadt, Germany}

\affiliation{Department of Physics and Astronomy, University of
  Tennessee, Knoxville, Tennessee 37996, USA}

\author{T. Papenbrock} 
\thanks{This manuscript has been authored by UT-Battelle, LLC under
  Contract No. DE-AC05-00OR22725 with the U.S. Department of
  Energy. The United States Government retains and the publisher, by
  accepting the article for publication, acknowledges that the United
  States Government retains a non-exclusive, paid-up, irrevocable,
  world-wide license to publish or reproduce the published form of
  this manuscript, or allow others to do so, for United States
  Government purposes. The Department of Energy will provide public
  access to these results of federally sponsored research in
  accordance with the DOE Public Access
  Plan. (http://energy.gov/downloads/doe-public-access-plan).}

\affiliation{Department of Physics and Astronomy, University of
  Tennessee, Knoxville, Tennessee 37996, USA}

\affiliation{Physics Division, Oak Ridge National Laboratory, Oak
  Ridge, Tennessee 37831, USA}

\date{\today}

\begin{abstract}
Heavy even-even nuclei exhibit low-energy collective excitations that
are separated in scale from the microscopic (fermion) degrees of
freedom. This separation of scale allows us to approach nuclear
vibrations within an effective field theory (EFT). In odd-mass nuclei
collective and single-particle properties compete at low energies, and
this makes their description more challenging. In this article we
describe odd-mass nuclei with ground-state spin $I=\sfrac{1}{2}$ by means
of an EFT that couples a fermion to the collective degrees of freedom
of an even-even core.  The EFT relates observables such as energy levels,
electric quadrupole ($E2$) transition strengths, and magnetic dipole
($M1$) moments of the odd-mass nucleus to those of its even-even neighbor,
and allows us to quantify theoretical uncertainties. For isotopes of
rhodium and silver the theoretical description is consistent with data
within experimental and theoretical uncertainties. Several testable
predictions are made.
\end{abstract}

\maketitle

\section{Introduction}
Collective modes such as rotations and vibrations are often the
lowest-lying excitations in heavy nuclei~\cite{bohr1975}, and these
phenomena can be understood in terms of collective
models~\cite{bohr1952, bohr1953, eisenberg1970-1, arima1975,
  arima1978, hess1980, matsuo1984, rowe2010} of the atomic nucleus.
In odd-mass nuclei, collective excitations compete with
single-particle excitations already at low energies. The well known
particle-rotor, particle-vibrator, and boson-fermion models couple the
odd fermion to the collective (boson) degrees of
freedom~\cite{deshalit1961,braunstein1962,iachello1979, iachello1980,
  vervier1982, vervier1983, vervier1984, vanisacker1984, jolie1985,
  frank1987, lampard1989, loiselet1989, maino1991}. While these models
successfully describe various aspects of odd-mass nuclei, it is
difficult to systematically improve them, or to give theoretical
uncertainties for the computed results.

In this paper, we want to re-examine odd-mass nuclei within an EFT
that couples a fermionic degree of freedom to the bosonic degrees of
freedom of the even-even nucleus. EFTs provide us with systematically
improvable approaches to nuclear interactions~\cite{vankolck1999,
  bedaque2002, epelbaum2009, machleidt2011,hammer2013}, clustering in
nuclei~\cite{bertulani2002, hammer2011, ryberg2014-1}, nuclear
rotations~\cite{papenbrock2011, zhang2013, papenbrock2014,
  papenbrock2015,coelloperez2015-1} and
vibrations~\cite{coelloperez2015-2}. They also allow us to quantify
theoretical
uncertainties~\cite{cacciari2011,bagnaschi2015,furnstahl2015-1,
  furnstahl2015-2}. This is an advantage over traditional models.
EFTs also allow us to derive relations between observables (opposed to
relations between model parameters and observables), and this makes
their application interesting even in cases where microscopic
approaches to nuclear collective phenomena are
available~\cite{boutachkov2002, paar2007, caprio2013, dytrych2013,
  caprio2015, jansen2015, stroberg2016}.

As EFTs are based on a separation of scales, we remind the reader
about the relevant low-energy scales in heavy nuclei.  In heavy
deformed even-even nuclei, rotational excitations (at about 0.1~MeV or
less) are separated in scale from vibrations (at about 0.8~MeV), which
in turn are separated from fermion excitations such as pair breaking
(at about 2-3~MeV). In heavy spherical even-even nuclei, vibrations
(at an energy $\omega\approx 0.6$~MeV) are lowest in energy and
separated from fermion excitations such as pair breaking at about
$\Lambda\approx~$2-3~MeV. In the recently proposed boson EFT for
nuclear vibrations~\cite{coelloperez2015-2}, the fermion energy
scale is the breakdown scale, and the ``small'' expansion parameter
is $\omega/\Lambda\approx 1/3$.

In this work we construct an EFT for odd-mass nuclei with spin
$\sfrac{1}{2}$ in their ground states by coupling an odd nucleon in a
$j=\sfrac{1}{2}$ orbital to the quadrupole degrees of freedom that
govern the collective vibrations of an even-even nucleus.  Based on a
power counting we systematically construct the Hamiltonian and
electromagnetic operators.  Another interesting aspect of this EFT
approach is the simultaneous description of the even-even and
neighboring odd-mass nuclei; consequently, observables in the
even-even nucleus are related to observables in the odd-mass system.
These relations can be confronted with experimental data. In this
work, we will compute $E2$ and $M1$ observables for odd-mass isotopes
of rhodium and silver. This is also interesting with view on recent
$g$ factor measurements in this region of the nuclear
chart~\cite{chamoli2011,stuchbery2016}.  The paper is organized as
follows. In Section~\ref{theory}, we present the EFT framework within
which the even-even/odd-mass nuclei will be described, establish a
power counting and describe energy spectra at next-to-next-to-leading
order. Sections~\ref{E2} and~\ref{M1} are dedicated to the study of
moments and transitions of the $E2$ and $M1$ operators,
respectively. In Section~\ref{discuss} we discuss the possible
extension of the EFT to the more complicated case posed by cadmium
isotopes. Finally, in Section~\ref{summary} we present our summary.

\section{Odd-mass vibrational nuclei}
\label{theory}
Certain even-even nuclei (such as isotopes of Cd, Ru, and Te) exhibit
low-energy states that resemble those of a five-dimensional quadrupole
oscillator. In these nuclei, the vibrational frequency $\omega\approx
0.6$~MeV is the energy scale of interest, and the picture of a
quadrupole vibrator breaks down at an energy $\Lambda\approx 2$-3~MeV,
i.e. around the three-phonon level. The breakdown scale $\Lambda$ is
associated with neglected microscopic (fermionic) degrees of freedom
and is of similar size as the pairing gap. Thus, $\omega\ll \Lambda$
holds, and this separation of scale has been exploited in
Ref.~\cite{coelloperez2015-2} to construct an EFT for nuclear
vibrations. 

The spectra of certain odd-mass neighbors of vibrational nuclei are
relatively simple and suggest that these result from coupling a
$j^\pi=\sfrac{1}{2}^-$ fermion to the even-even nucleus. Examples we
consider in this paper are $^{99,101,103}$Rh and
$^{105,107,109,111}$Ag as a proton coupled to $^{98,100,102}$Ru
and $^{104,106,108,110}$Pd, respectively, or $^{107,109,111}$Ag
as a proton-hole in $^{108,110,112}$Cd. These cases are particularly
simple because one deals with a $j^{\pi}=\sfrac{1}{2}^{-}$ degree of
freedom. We note here that the odd-mass nuclei considered in this work
also exhibit very low-lying (100~keV or less) states with positive parity.
As a single fermion cannot undergo parity-changing transitions, the
positive-parity states can be neglected in the description of low-lying
negative-parity states in the odd-mass nuclei.

Could one also attempt to describe, for instance, $^{108,110,112}$Cd
in terms of two protons added to $^{106,108,110}$Pd, respectively? In
such an EFT approach, the low-lying positive-parity states of
$^{107,109,111}$Ag would also need to enter the description. The
calculation would be non-perturbative (because of the near degeneracy
of states with positive and negative parities in the odd-mass
nucleus), and a significant number of fermionic two-body-matrix
elements would enter as low energy constants (LECs). It is thus
unclear whether such an EFT approach would be profitable.

\subsection{Hamiltonian}

Before we turn to the odd-mass nuclei, we briefly review some aspects
of the EFT for nuclear vibrations in even-even
nuclei~\cite{coelloperez2015-2}. The relevant degrees of freedom are
quadrupole operators $d^{\dagger}_{\mu}$ and $d_{\mu}$ with
$\mu=-2,-1,...,2$ that create and annihilate a phonon,
respectively. They fulfill the usual boson commutation relations
\begin{equation}
\left[d_{\mu},d^{\dagger}_{\nu}\right]=\delta_{\mu\nu} . 
\label{quadcomm}
\end{equation}
We note that $d^{\dagger}_{\mu}$ and 
\begin{equation}
\tilde{d}_{\mu} = (-1)^{\mu}d_{-\mu} 
\end{equation} 
are spherical tensors of rank two. The angular momentum operator for
the quadrupole degrees of freedom is the vector
\begin{equation}
\hat{\mathbf{J}} = \sqrt{10}\left(d^{\dagger}\otimes\tilde{d}\right)^{(1)} .  
\label{Jbos}
\end{equation}
We recall that the coupling of the spherical tensors $\mathcal{M}^{(m)}$ and
$\mathcal{N}^{(n)}$ of ranks $m$ and $n$, respectively, to a spherical tensor
$\mathcal{K}^{(k)}$ of rank $k$ is denoted as
\begin{equation}
\mathcal{K}^{(k)} = \left(\mathcal{M}^{(m)}\otimes\mathcal{N}^{(n)}\right)^{(k)} , 
\end{equation}
and the corresponding components 
\begin{equation}
\mathcal{K}_{\kappa}^{(k)} = \sum_{\mu \nu} C_{m \mu n \nu }^{k \kappa} \mathcal{M}_{\mu}^{(m)} \mathcal{N}_{\nu}^{(n)}  
\end{equation}
are given in terms of the Clebsch-Gordan coefficients $C_{m \mu n \nu
}^{k \kappa}$ that couple the angular momenta $m$ and $n$ to spin
$k$~\cite{varshalovich1988}. Similarly, the scalar product of two
spherical tensors $\mathcal{M}^{(I)}$ and $\mathcal{N}^{(I)}$ of the
same rank $I$ is~\cite{varshalovich1988}
\begin{eqnarray}
\mathcal{M}^{(I)} \cdot \mathcal{N}^{(I)} &=& \sqrt{2I+1} \left(\mathcal{M}^{(I)} \otimes \mathcal{N}^{(I)}\right)^{(0)}
\end{eqnarray}

The boson Hamiltonian is
\begin{eqnarray}
\label{Hb}
\hat{H}_{\rm b} &=& \omega_1\hat{N} + g_{N}\hat{N}^{2} + g_{v}\hat{\Lambda}^{2} + g_{J}\hat{J}^{2} . 
\end{eqnarray} 
Here, 
\begin{equation}
\hat{N} \equiv d^{\dagger}\cdot\tilde{d}
\end{equation}
and
\begin{equation}
\hat{\Lambda}^{2} \equiv -\left(d^{\dagger}\cdot d^{\dagger}\right)\left(\tilde{d}\cdot\tilde{d}\right)+\hat{N}^{2}-3\hat{N}
\end{equation}
are the boson number operator and the SO(5) equivalent of the SO(3)
angular momentum squared operator $\hat{J}^{2}$. For more details on
the later operator and its eigenvalues see, for example,
Ref.~\cite{rowe2010}. The first term on the right-hand side of
Eq.~(\ref{Hb}) is of order $\omega$. This leading order (LO) term is
the Hamiltonian of a five-dimensional harmonic oscillator. The
remaining terms in the Hamiltonian~(\ref{Hb}) account for finer
details at order $\omega^3/\Lambda^2$. These corrections introduce
anharmonicities. The power counting of the EFT is in powers of the small
parameter $\omega/\Lambda$. For details, we refer the reader to
Ref.~\cite{coelloperez2015-2}.

The fermion is described in terms of fermion creation and
annihilation operators $a^{\dagger}_{\nu}$ and $a_{\nu}$
respectively, that fulfill the usual anticommutation relations
\begin{equation}
\left\{a_{\mu},a^{\dagger}_{\nu}\right\}=\delta_{\mu\nu}.
\label{fercomm}
\end{equation}
In most of this paper, $\nu=-\sfrac{1}{2}, \sfrac{1}{2}$.
The corresponding angular momentum operator is 
\begin{equation}
\hat{\mathbf{j}} = {1\over\sqrt{2}}\left(a^{\dagger}\otimes\tilde{a}\right)^{(1)},
\end{equation}
and the fermion number operator is
\begin{equation}
\hat{n} = a^{\dagger}\cdot\tilde{a}. 
\end{equation}
Here, we used the spherical rank-$\sfrac{1}{2}$ tensor $\tilde{a}$ with components
\begin{equation}
\tilde{a}_{\nu} \equiv (-1)^{j+\nu}a_{-\nu} .
\end{equation}

The fermion Hamiltonian 
\begin{equation}
\label{Hf}
\hat{H}_{\rm f} = -S \hat{n} - \Delta \hat{n}(\hat{n}-1)   
\end{equation}
consists of a one-body term and a two-body term. We note that the term
$\hat{n}(\hat{n}-1)$ is the unique two-body interaction for
spin-$\sfrac{1}{2}$ fermions restricted to a single $j^{\pi}=
\sfrac{1}{2}^{+}$ shell.  We do not need to consider other Hamiltonian
terms such as $\hat{\mathbf{j}}^2\propto \hat{n}(2-\hat{n})$ or
$\hat{n}^2$ because these are linear combinations of the terms
already included in the Hamiltonian~(\ref{Hf}).

The Hamiltonian~(\ref{Hf}) is not the Hamiltonian of free fermions but
rather captures the interactions between fermions and the ground state
of the vibrating core. Let us discuss the energy scales $S$ and
$\Delta$.  For a particle (hole) added to the even-even vibrator,
$S\approx 8$~MeV ($S\approx -8$~MeV) is of order of the separation
energy, while $\Delta\approx 2$~MeV is of the order of a paring
gap. The attractive interaction between two nucleons (with isospin
one) fail to bind the pair in vacuum but yields a bound state with
energy $\Delta$ when coupled to the core. We note that
$\Delta\sim\Lambda$, as pairing effects are one source of the
breakdown scale of the even-even nucleus.

The interaction between boson and fermion degrees of freedom is
most interesting. Two-body terms of the structure
$\hat{\mathbf{J}}\cdot\hat{\mathbf{j}}$ and $\hat{N}\hat{n}$ couple
phonons to fermions. Here, the first term could be referred to as a
``Coriolis'' interaction, because it couples the spin of the fermion
to the spin of the core. In addition to these interactions
there are three-body terms of the forms $\hat{N}^2\hat{n}$,
$\hat{\mathbf{J}}^2\hat{n}$, and $\hat{N}\hat{n}(\hat{n}-1)$. Here,
the first two three-body terms involve the annihilation and creation
of two phonons and are suppressed in comparison to the three-body term
involving only one phonon. Thus, the leading-order interactions between
phonons and fermion degrees of freedom are
\begin{equation}
\label{Hbf}
H_{\rm b-f} = g_{Jj}\hat{\mathbf{J}}\cdot\hat{\mathbf{j}} + 
\omega_2 \hat{N}\hat{n} + \omega_3\hat{N}\hat{n}(\hat{n}-1).
\end{equation}
We note that the three-body term $\omega_3\hat{N}\hat{n}(\hat{n}-1)$
is only active when two fermions are coupled to the vibrating core.

Let us attempt to establish a power counting for operators involving
fermion degrees of freedom. For an operator $\hat{O}_{n}$ with
$2n$ fermion operators, we propose its matrix elements to scale as
\begin{equation}
\langle \hat{O}_{n} \rangle \sim \langle \hat{O}_{n-1} \rangle
\frac{\omega}{\Lambda}.
\label{nbodypower}
\end{equation}
This scaling is consistent with the energy spectra of the systems we
study. Thus, the terms involving one fermion in the interaction
Hamiltonian~(\ref{Hbf}) scale as $\omega^{2}/\Lambda$.

Putting everything together, and
restricting ourselves to a single fermion, we arrive at the Hamiltonian
\begin{eqnarray}
\label{ham}
H &=& H_{\rm b} + H_{\rm f} + H_{\rm b-f} \nonumber\\
  &=& -S \hat{n} + H_{\rm LO} + H_{\rm NLO} + H_{\rm NNLO} , 
\end{eqnarray}
with
\begin{equation}
\label{hlo}
H_{\rm LO} \equiv \omega_1\hat{N},
\end{equation}
\begin{equation}
\label{hnlo}
H_{\rm NLO} \equiv g_{Jj}\hat{\mathbf{J}}\cdot\hat{\mathbf{j}}
+ \omega_2 \hat{N}\hat{n}
\end{equation}
and
\begin{equation}
\label{hnnlo}
H_{\rm NNLO} \equiv g_{N}\hat{N}^{2} + g_{v}\hat{\Lambda}^{2}
+ g_{J}\hat{J}^{2}.
\end{equation}
While the term $-S\hat{n}$ in Eq.~(\ref{ham}) sets the overall binding
with respect to the ground-state of the vibrating core, it does not
contain any spectroscopic information. We will therefore neglect this
term in what follows.  The LO Hamiltonian~(\ref{hlo}) is that of a
harmonic quadrupole vibrator, and energies are of the order
$\omega$. Higher-order contributions to the Hamiltonian are most
interesting. The NLO Hamiltonian~(\ref{hnlo}) accounts for effects
introduced by the phonon-fermion couplings. We note that the size of
the boson-fermion interaction cannot be determined on theoretical
grounds but must rather be based on data.  The empirical inspection of
spectra suggests that these phonon-fermion couplings are a fraction of
the scale $\omega$. We approximate this scale as order
$\omega^2/\Lambda$ and thereby avoid the introduction of a new small
parameter. Because of this perturbative coupling we can associate
low-lying states in certain odd-mass nuclei with the spectra in the
neighboring even-even nuclei. The NNLO Hamiltonian~(\ref{hnnlo})
involves phonon-phonon interactions that account for anharmonicities
in the even-even nucleus. We remind the reader that these terms are of
order $\omega^3/\Lambda^2$ and have been discussed in detail in
Ref.~\cite{coelloperez2015-2}.
  
Let us discuss the Hilbert space.  The states of the odd nucleus are
products of the boson quadrupole states and fermion states of the
$j=\sfrac{1}{2}$ orbital.  As usual, the vacuum $|0\rangle$ fulfills
\begin{equation}
d_{\mu}|0\rangle = 0 = a_\nu|0\rangle .
\end{equation}
The boson states of the quadrupole vibrator are created from the
vacuum by the successive application of quadrupole creation
operators. These states are denoted as
\begin{equation}
|N\alpha v J\mu \rangle .
\end{equation}
Here $N$ is the number of phonons, $v$ is the seniority, $J$ and $\mu$
are the angular momentum and its projection onto the $z$-axis,
respectively, while $\alpha$ represents an additional quantum
number. This quantum number is only needed above the two-phonon level
and therefor not needed for the low-energy physics we are interested
in. We will omit it in what follows. For details on the construction
of these states we refer the reader to
Ref.~\cite{rowe2010}. The single-fermion states are
\begin{equation}
|\tfrac{1}{2}\nu\rangle \equiv a^\dagger_\nu|0\rangle .
\end{equation}
Normalized states of the odd-mass nucleus with total spin $I$ and projection $M$ are 
\begin{eqnarray}
\label{states}
\lefteqn{|IM;N\alpha vJ;\tfrac{1}{2}\rangle \equiv  \left(|N\alpha vJ\rangle \otimes |\tfrac{1}{2}\rangle\right)^{(I)}_{M}}\nonumber\\
&& = \sum_{\mu\nu}C_{J\mu \frac{1}{2}\nu}^{IM} |N\alpha vJ\mu\rangle |\tfrac{1}{2}\nu\rangle .
\end{eqnarray}

\begin{figure}[htb!]
\centering
\includegraphics[width=0.485\textwidth]{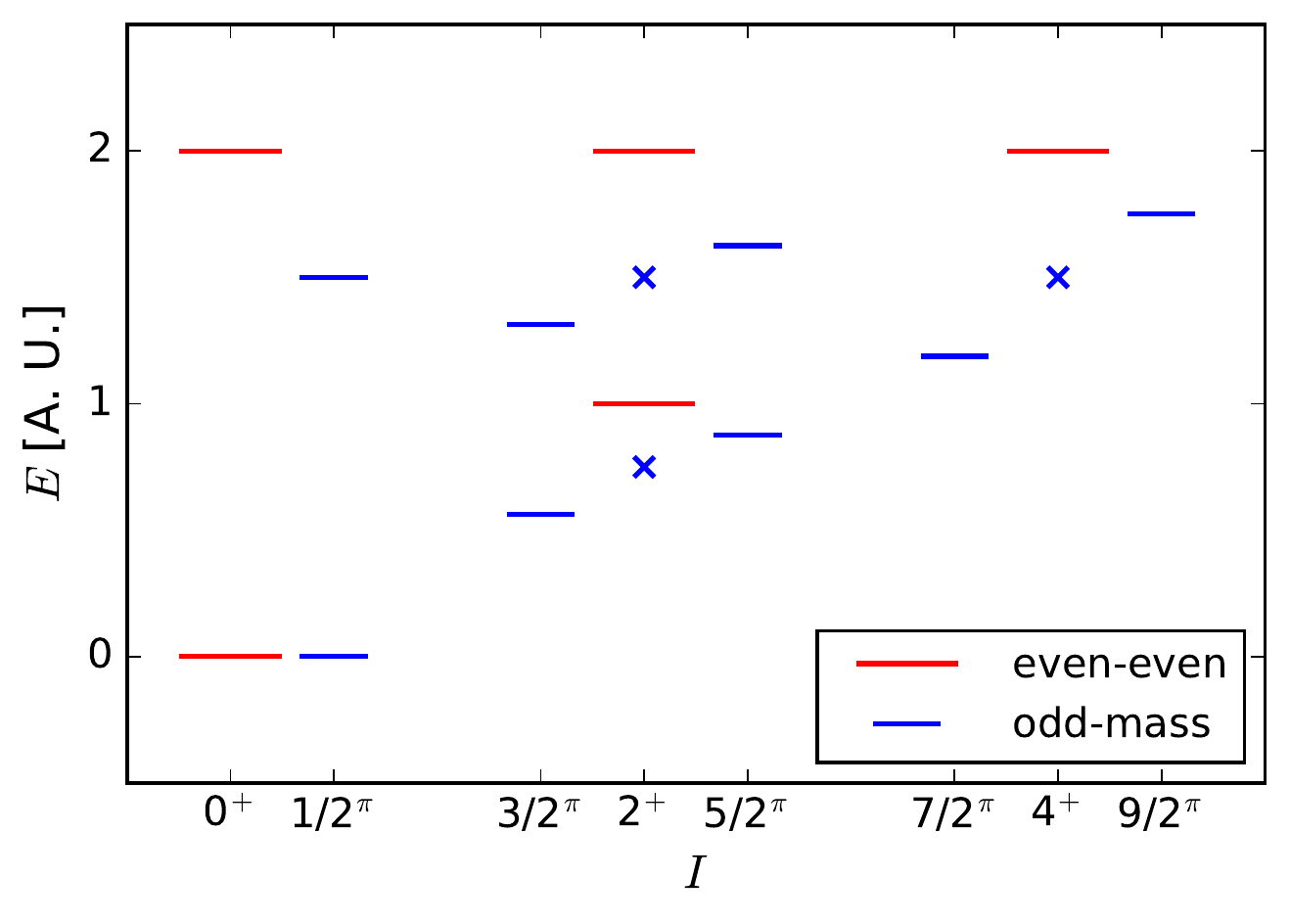}
\caption{(Color online) NLO spectrum for the fermion in a
  $j=\sfrac{1}{2}$ orbital coupled to a quadrupole vibrator up to the
  two-phonon level in arbitrary units. The states labeled as
  $I^{\pi}$, with $\pi$ being the parity, are displayed as long red
  and short blue lines for even-even and odd-mass nuclei,
  respectively. The ``centers of gravity'' of the $I=J\pm j$ odd-mass
  states are shown as blue crosses.}
\label{spectrum}
\end{figure}

The Hamiltonian~(\ref{ham}) is diagonal in the basis
states~(\ref{states}) with eigenvalues
\begin{equation}
\label{NLOE}
E = E_{\rm LO} + E_{\rm NLO} + E_{\rm NNLO},
\end{equation}
with
\begin{equation}
\label{elo}
E_{\rm LO} = \omega_{1}N,
\end{equation}
\begin{equation}
\label{enlo}
E_{\rm NLO} = \omega_2 N n + \frac{g_{Jj}}{2} \left[I(I+1)-J(J+1)-\frac{3}{4}\right]
\end{equation}
and
\begin{equation}
\label{ennlo}
E_{\rm NNLO} = g_{N}N^{2} + g_{v}v(v+3) + g_{J}J(J+1).
\end{equation}
We remind the reader that we neglected the separation energy $S$,
i.e., the ground-state energies of the even-even nucleus and of the
odd-mass nucleus are set to zero.  Figure~\ref{spectrum} shows a
schematic plot of the NLO energy spectrum~(\ref{NLOE}) up to the
two-phonon level. States are labeled by their spin and
parity. Even-even states, shown as long red lines, have integer spins
and positive parity.  Odd-mass states, shown as short blue lines, have
half-integer spins and the parity of the fermion's orbital. (Odd-mass
states considered in what follows all have negative parities.)
Energies are chosen in units of $\omega_1$, and the LECs $\omega_2$
and $g_Jj$ are small fractions of this LEC.  We see how the term
proportional to $\omega_2$ shifts the energies while the term
proportional to $g_{Jj}$ splits even-even states with finite spins
into doublets in the odd-mass neighbor. The ``centers of gravity''
from the shift are shown as crosses in Fig.~\ref{spectrum}.

\subsection{Uncertainty quantification}
EFTs provide us with the opportunity to quantify theoretical
uncertainties. While the power counting allows one to estimate
uncertainties in EFTs, quantified uncertainties result from (testable)
assumptions one makes about the distribution of the LECs
~\cite{furnstahl2015-2} in form of priors. Employing Bayesian
statistics (and marginalizing) over unknown parameters included in these
priors yields degree-of-belief (DOB) intervals with a statistical
meaning. In this section, we closely follow
Ref.~\cite{coelloperez2015-2} and chose log-normal priors for the LECs'
distribution functions. 

The energies of the states below the breakdown scale can be written as
an expansion of the form
\begin{equation}
  E(I^{\pi}) = \omega_1\sum\limits_{i}^{\infty}c_{i}(I^{\pi})\varepsilon^{i}
\end{equation}
with
\begin{equation}
  \varepsilon\equiv N{\omega_1\over \Lambda}.
\end{equation}
In our case
\begin{equation}
\label{vareps}
{\omega_{1}\over\Lambda}\approx {1\over 3}.
\end{equation}
If the expansion is truncated at order
$\mathcal{O}(\varepsilon^{2})$, a comparison with the NNLO
spectrum~(\ref{NLOE}) allows us to identify
\begin{equation}
c_{0}(I^{\pi})\equiv \frac{E_{\rm LO}(I^{\pi})}{\omega_{1}},
\end{equation}
\begin{equation}
c_{1}(I^{\pi})\equiv \frac{E_{\rm NLO}(I^{\pi})}{\varepsilon \omega_{1}}
\end{equation}
and
\begin{equation}
c_{2}(I^{\pi})\equiv \frac{E_{\rm NNLO}(I^{\pi})}{\varepsilon^{2}\omega_1}
\end{equation}
From the power counting one expects these coefficients to be of order
$\mathcal{O}(1)$.

\begin{figure}
\centering
\includegraphics[width=0.49\textwidth]{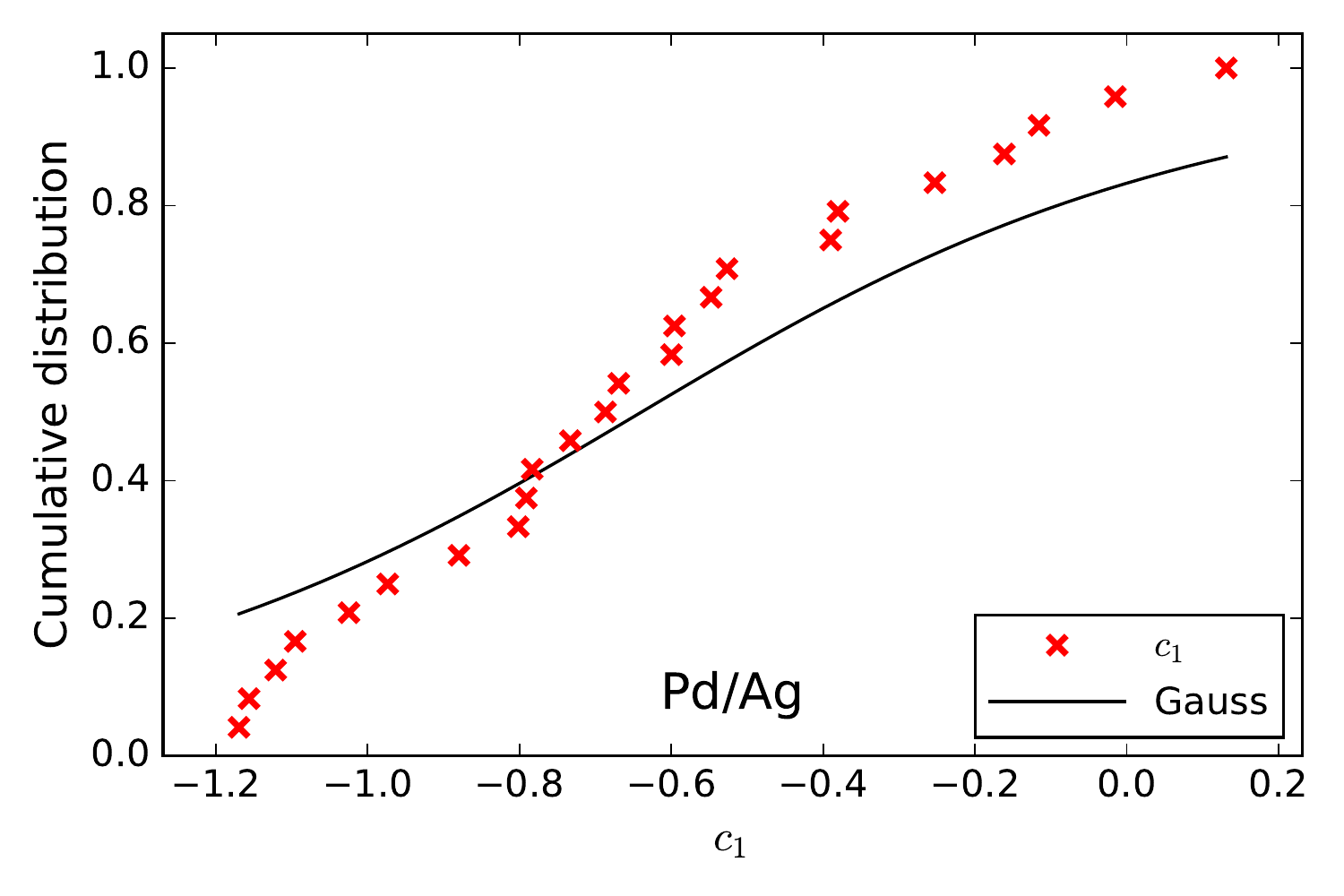}
\includegraphics[width=0.49\textwidth]{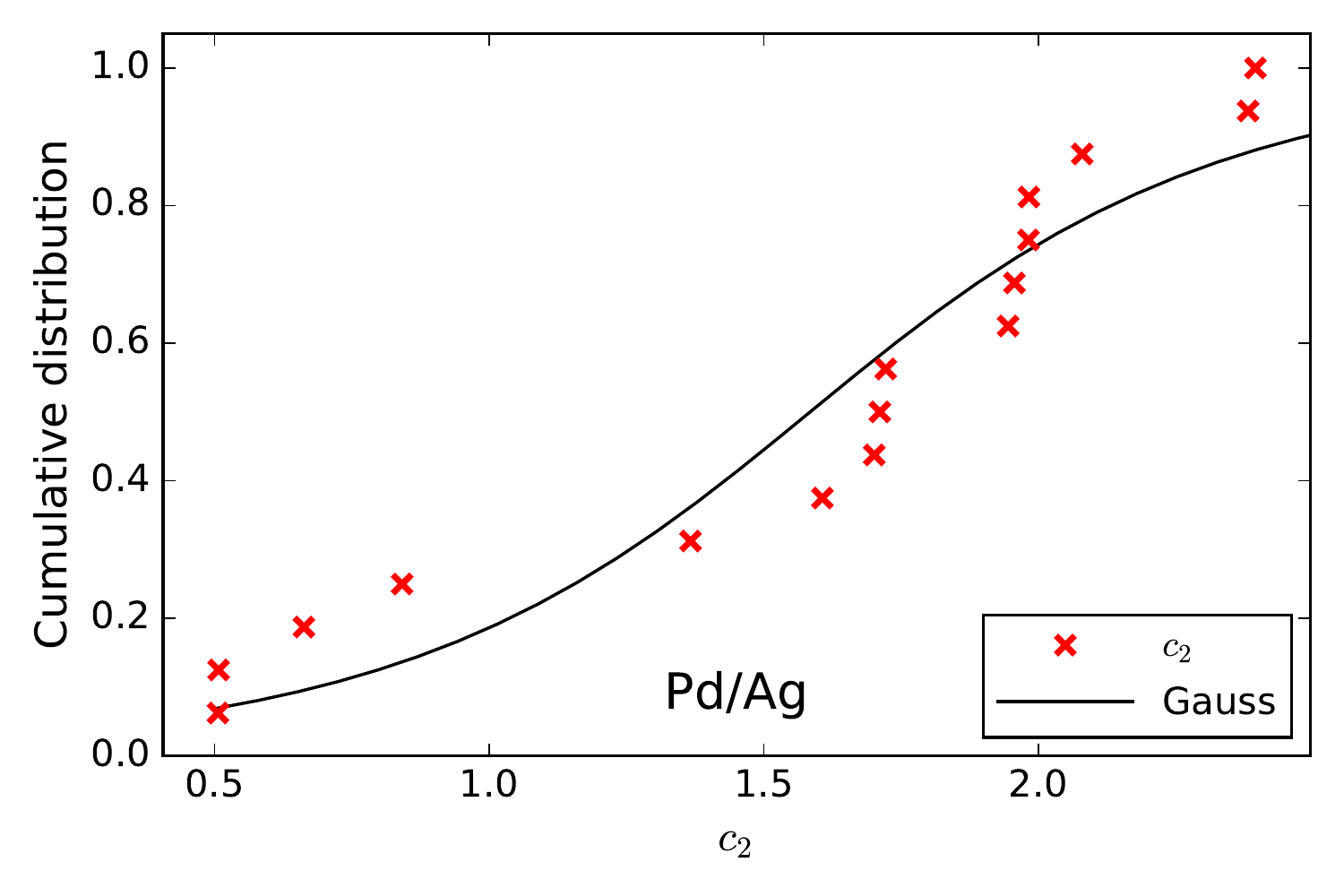}
\caption{(Color online) Cumulative distributions of the $c_{1}$ (top)
  and $c_{2}$ (bottom) coefficients for the energies of states below
  breakdown in an ensemble containing the data of all studied Pd and
  Ag nuclei. These distributions, centered at $\mu_{1}$ and $\mu_{2}$,
  are approximated by Gaussian priors (shown as lines).}
\label{cumulative}
\end{figure}

Figure~\ref{cumulative} shows the cumulative distributions of the
$c_{1}$ and $c_{2}$ coefficients for the energies of states below the
breakdown scale in an ensemble containing the data of all studied Pd
and Ag nuclei.
These distributions, with means $\mu_{1}$ and $\mu_{2}$, respectively,
can be approximated by the Gaussian prior
\begin{equation}
{\rm pr}^{(\rm G)}(\tilde{c}_{i}|c)=\frac{1}{\sqrt{2\pi}sc}e^{-\frac{\tilde{c}_{i}^{2}}{2s^{2}c^{2}}} \quad {\rm with} \quad s=\frac{2}{3}
\label{Gprior}
\end{equation}
for the expansion coefficient $c_{i}=\tilde{c}_{i}+\mu_{i}$. Here,
$\mu_{i}\equiv\overline{c_i}$ is the mean value of the $c_i$. The
parameter $c$, associated with the width of the distribution, is not
taken from Fig.~\ref{cumulative}. Instead, we make the assumption that
$c$ is log-normal distributed according to
\begin{equation}
{\rm pr}(c)=\frac{1}{\sqrt{2\pi}\sigma c}e^{-\frac{\log^{2}{c}}{2\sigma^{2}}} .
\label{lognor}
\end{equation}
The log normal distribution is consistent with the EFT expectation
that LECs are of natural size, i.e. that the coefficient $c$ is of
order one~\cite{cacciari2011}. Given the priors~(\ref{Gprior})
and~(\ref{lognor}), one calculates the probability distribution
function (PDF) for $c_{i}$ by marginalizing over the parameter
$c$ and finds
\begin{equation}
p(c_{i}-\mu)=\int\limits_{0}^{\infty}dc {\rm pr}^{(\rm G)}
(c_{i}-\mu_{i}|c){\rm pr}(c).
\label{cprior}
\end{equation}
The cumulative distribution for $c_{i}$, denoted by
${\rm CDF}(c_{i})$, is then given in terms of the PDF~(\ref{cprior}) by
\begin{equation}
{\rm CDF}(c_{i})=\int\limits_{-\infty}^{c_{i}}dx p(x-\mu_{i}).
\end{equation}

Bayesian methods can be employed to quantify the uncertainties
associated to the energies~\cite{cacciari2011, furnstahl2015-2}
at any order. From the EFT expansion for an observable
\begin{equation}
X=X_{0}\sum\limits_{i}^{\infty}c_{i}\varepsilon^{i},
\label{EFTobs}
\end{equation}
it is clear that an order-$k$ calculation has a normalized uncertainty
that can be approximately written as
\begin{equation}
\Delta^{(k)} = \sum\limits_{i=k+1}^{k+M}c_{i}\varepsilon^{i}.
\end{equation}

The PDF for the normalized uncertainty can be calculated from the
priors for the expansion coefficients~(\ref{Gprior}) and the width
parameter~(\ref{lognor}) via Bayesian methods. We employed the
expressions given in Ref~\cite{coelloperez2015-2} to calculate the PDF
for the normalized uncertainty given the known coefficients, denoted
by $p(\Delta|c_{0},\ldots, c_{k})$, within the next-term
approximation, that is, setting $M=1$.

Given $p(\Delta|c_{0},\ldots, c_{k})$, the DOB of the interval
$[\alpha,\beta]$ is defined by
\begin{equation}
{\rm DOB}(\alpha,\beta)=\int\limits_{\alpha}^{\beta}d\Delta p(\Delta|c_{0},\ldots, c_{k}).
\end{equation}
We employ an interval of the form $[-\delta,\delta]$ with ${\rm
  DOB}(-\delta,\delta)=0.68$ to quantify the uncertainty $\Delta X^{(k)}$
associated to the order-$k$ calculation for $X$ as
\begin{equation}
\Delta X^{(k)}\equiv X_{0}\delta.
\end{equation}
Statistically, one expects 68\% of the experimental data to fall
within the theoretical uncertainty quantified by this DOB interval.

\subsection{Spectra}
We need to adjust the LECs of our EFT to data from an even-even and an
odd-mass nucleus simultaneously.  The spectra of such an
even-even/odd-mass system must resemble Eq.~(\ref{NLOE}),
schematically shown in Figure~\ref{spectrum}.  We recall that the EFT
does not distinguish between a fermion particle or a fermion hole.
This allows us to describe the isotopes $^{107,109,111}$Ag as a proton
coupled to $^{106,108,110}$Pd or as a proton hole coupled to
$^{108,110,112}$Cd.  Assuming the validity of our EFT approach, both
descriptions should agree within theoretical uncertainties.

Table~\ref{LOELECs} lists the LECs for the systems studied in this
work. Odd-mass nuclei in these systems have $I^{\pi}=
\sfrac{1}{2}^{-}$ ground states. The LECs were fitted employing data
for the energies of states identified as one- or two-phonon
levels. Most of the states employed in the fit have definite
assignments of spins and parities. For states with tentative spins, we
made the following assignments: $I^{\pi}=\sfrac{3}{2}^{-}$ for the
state at 410.9~keV in $^{99}{\rm Rh}$; $I^{\pi}=\sfrac{3}{2}^{-}$ and
$I^{\pi}= \sfrac{7}{2}^{-}$ for the states at 305.4 and 851.3~keV in
$^{101}{\rm Rh}$, respectively; $I^{\pi}=\sfrac{7}{2}^{-}$ for the
state at 973.3~keV in $^{107}{\rm Ag}$. These assignments were based
on the decay patterns from these states to other phonon states and
they agree with tentative spin assignments. The data were taken from
Refs.~\cite{defrenne1996, blachot1998, blachot2000, singh2003,
  defrenne2005, blachot2006, blachot2007, blachot2008, defrenne2008,
  singh2008, blachot2009, defrenne2009-1, defrenne2009-2, browne2011,
  gurdal2012}.

\begin{table}
\caption{LECs in keV employed to generate the NNLO spectra of selected
  even-even/odd-mass systems studied in this work.}  \centering
\begin{tabular}{cd{3.1}d{4.1}d{2.1}d{3.1}d{3.1}d{2.1}}
\hline\hline
System & \multicolumn{1}{c}{$\omega_1$} &
\multicolumn{1}{c}{$\omega_2$} & \multicolumn{1}{c}{$g_{Jj}$} &
\multicolumn{1}{c}{$g_{N}$} & \multicolumn{1}{c}{$g_{v}$} &
\multicolumn{1}{c}{$g_{J}$} \\
\hline
$^{98}{\rm Ru}/^{99}{\rm Rh}$ & 570.4 & -231.3 & 6.8 & 45.3 & 9.3 & -0.1 \\
$^{100}{\rm Ru}/^{101}{\rm Rh}$ & 376.6 & -204.0 & 19.9 & 94.3 & 27.2 & -6.7 \\
$^{102}{\rm Ru}/^{103}{\rm Rh}$ & 341.3 & -141.9 & 24.9 & 83.2 & 9.3 & 2.2 \\
$^{104}{\rm Pd}/^{105}{\rm Ag}$ & 439.3 & -157.1 & 34.5 & 115.8 & -8.9 & 6.1 \\
$^{106}{\rm Pd}/^{107}{\rm Ag}$ & 382.7 & -127.7 & 39.3 & 92.1 & -6.5 & 10.5 \\
$^{108}{\rm Cd}/^{107}{\rm Ag}$ & 576.1 & -249.9 & 39.3 & 142.1 & -30.6 & 6.2 \\
$^{108}{\rm Pd}/^{109}{\rm Ag}$ & 407.6 & -59.1 & 41.6 & 100.1 & -37.2 & 12.4 \\
$^{110}{\rm Cd}/^{109}{\rm Ag}$ & 606.5 & -283.4 & 41.6 & 109.2 & -32.2 & 11.7 \\
$^{110}{\rm Pd}/^{111}{\rm Ag}$ & 334.3 & -22.4 & 40.7 & 92.1 & -32.0 & 12.5 \\
$^{112}{\rm Cd}/^{111}{\rm Ag}$ & 543.9 & -265.6 & 40.7 & 82.7 & -21.1 & 12.4 \\
\hline\hline
\end{tabular}
\label{LOELECs}
\end{table}
\begin{figure}[t!]
\centering
\includegraphics[width=0.49\textwidth]{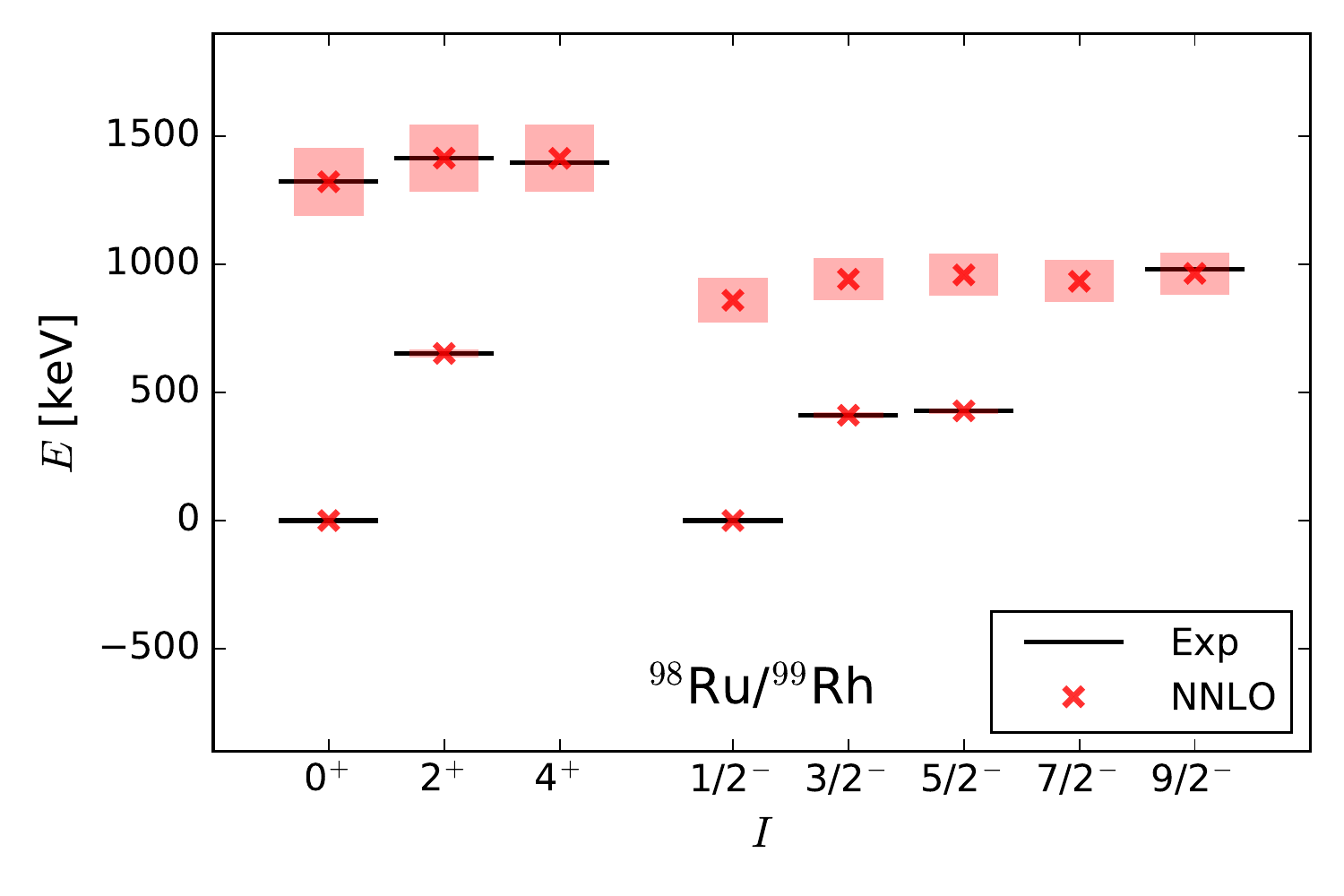}
\includegraphics[width=0.49\textwidth]{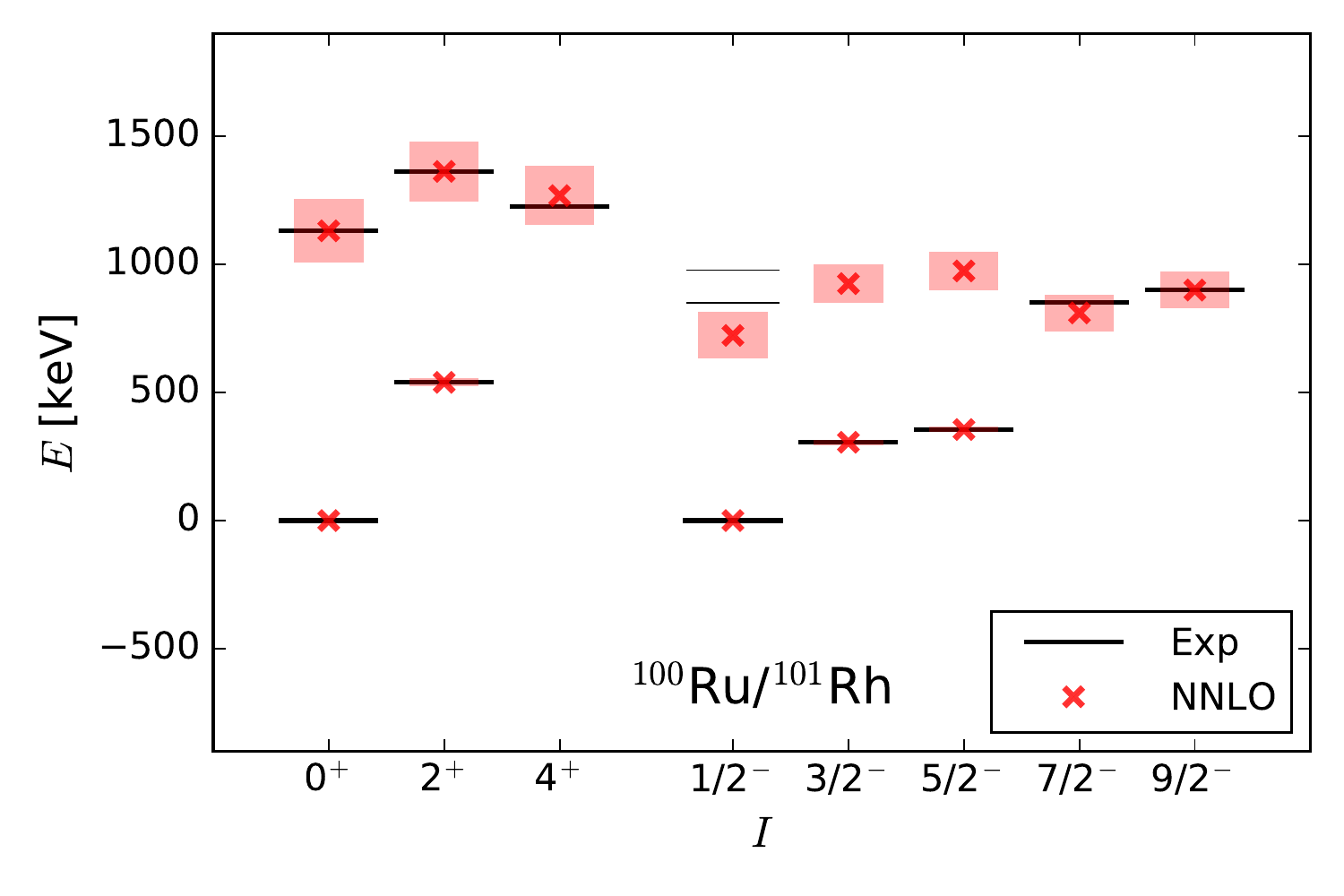}
\includegraphics[width=0.49\textwidth]{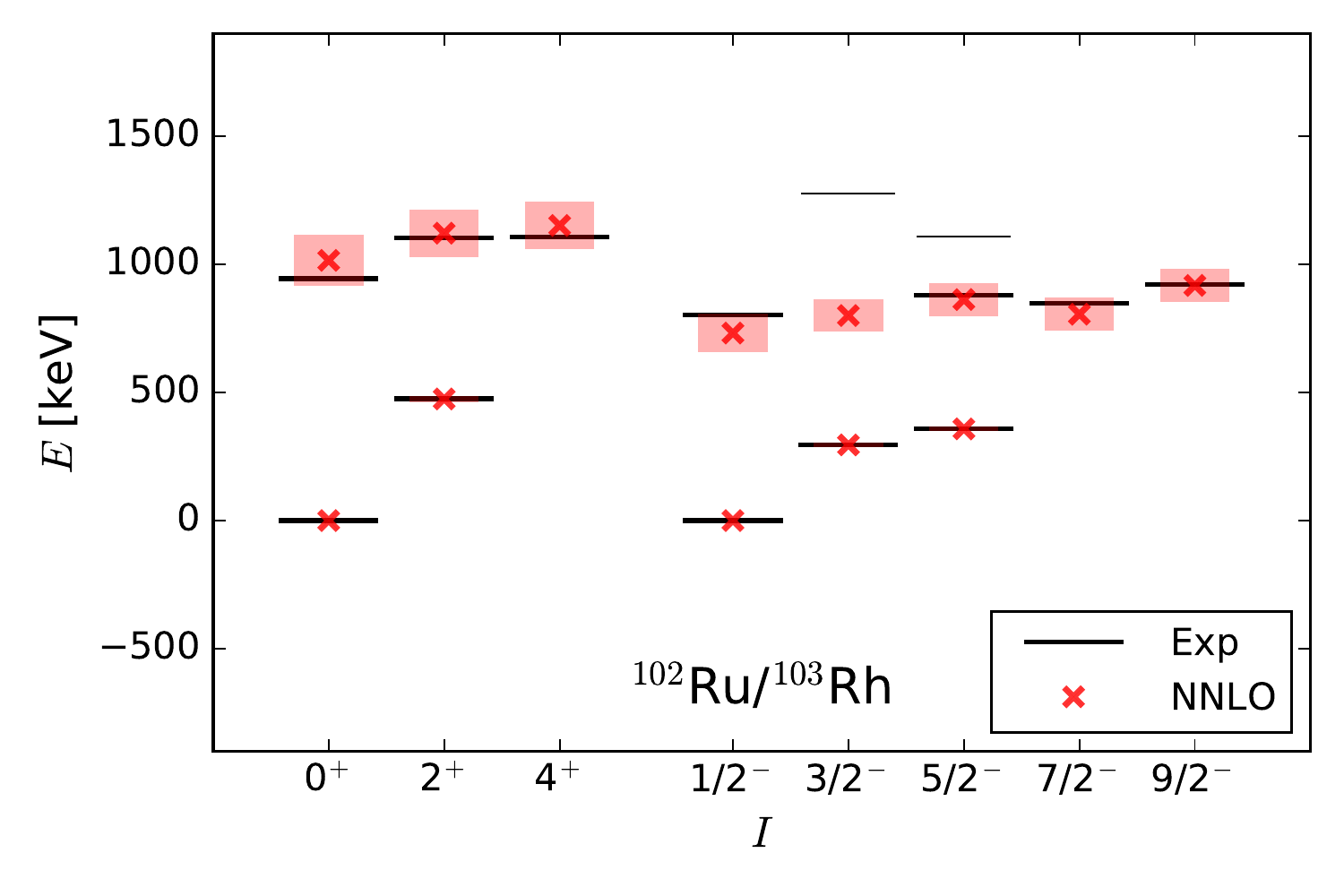}
\caption{(Color online) NNLO energy spectra of Ru/Rh systems. Rh is
  described as a proton in a $j^{\pi}=\sfrac{1}{2}^{-}$ orbital
  coupled to a Ru core.  Thick black lines denote states employed to
  fit the LECs while thin black lines denote states with a definitely
  known spin or a single tentative spin-parity assignment. Red crosses
  and shaded areas denote theoretical predictions and uncertainties,
  respectively.}
\label{rurh}
\end{figure}

\begin{figure*}[htb!]
\centering
\includegraphics[width=0.49\textwidth]{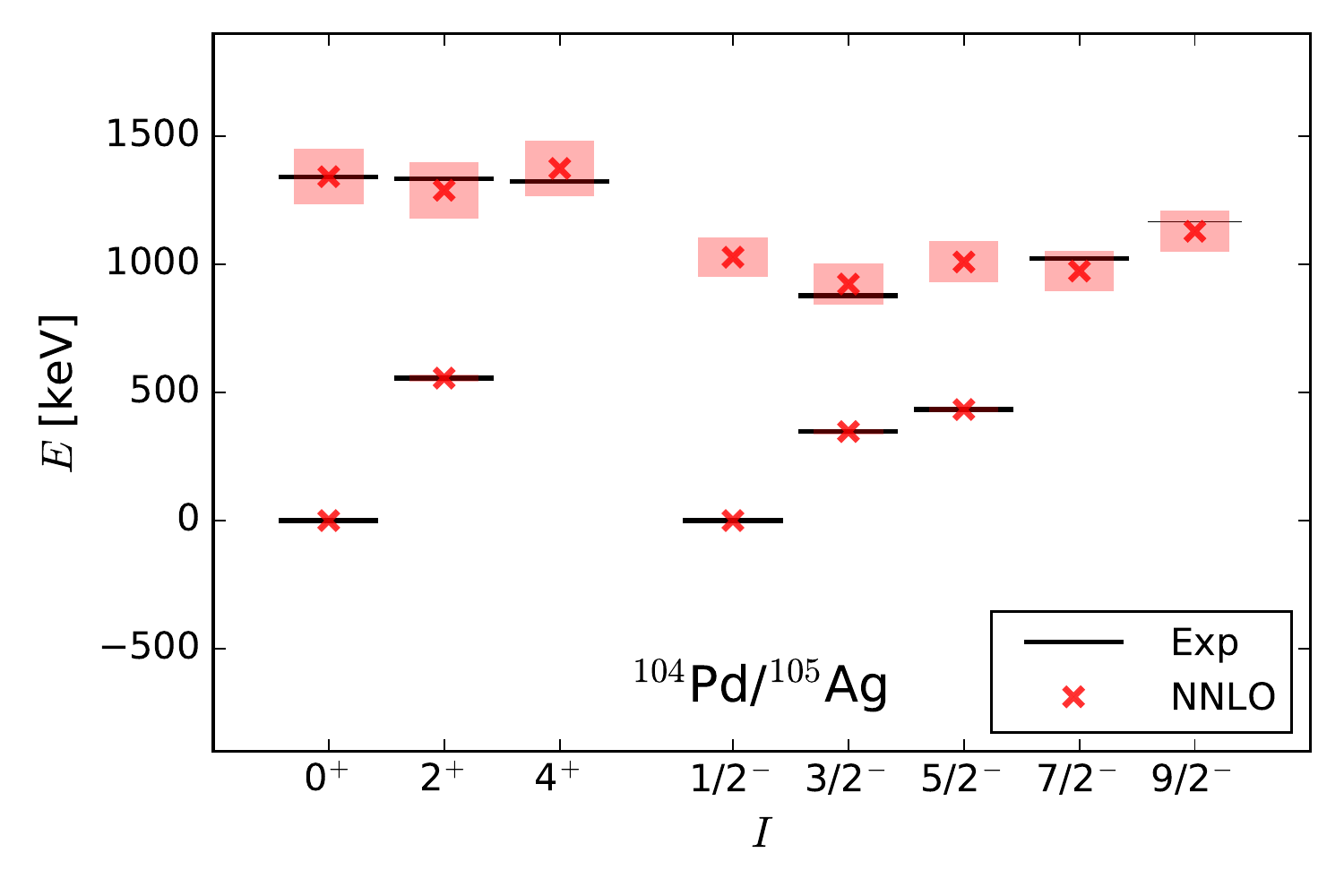}
\includegraphics[width=0.49\textwidth]{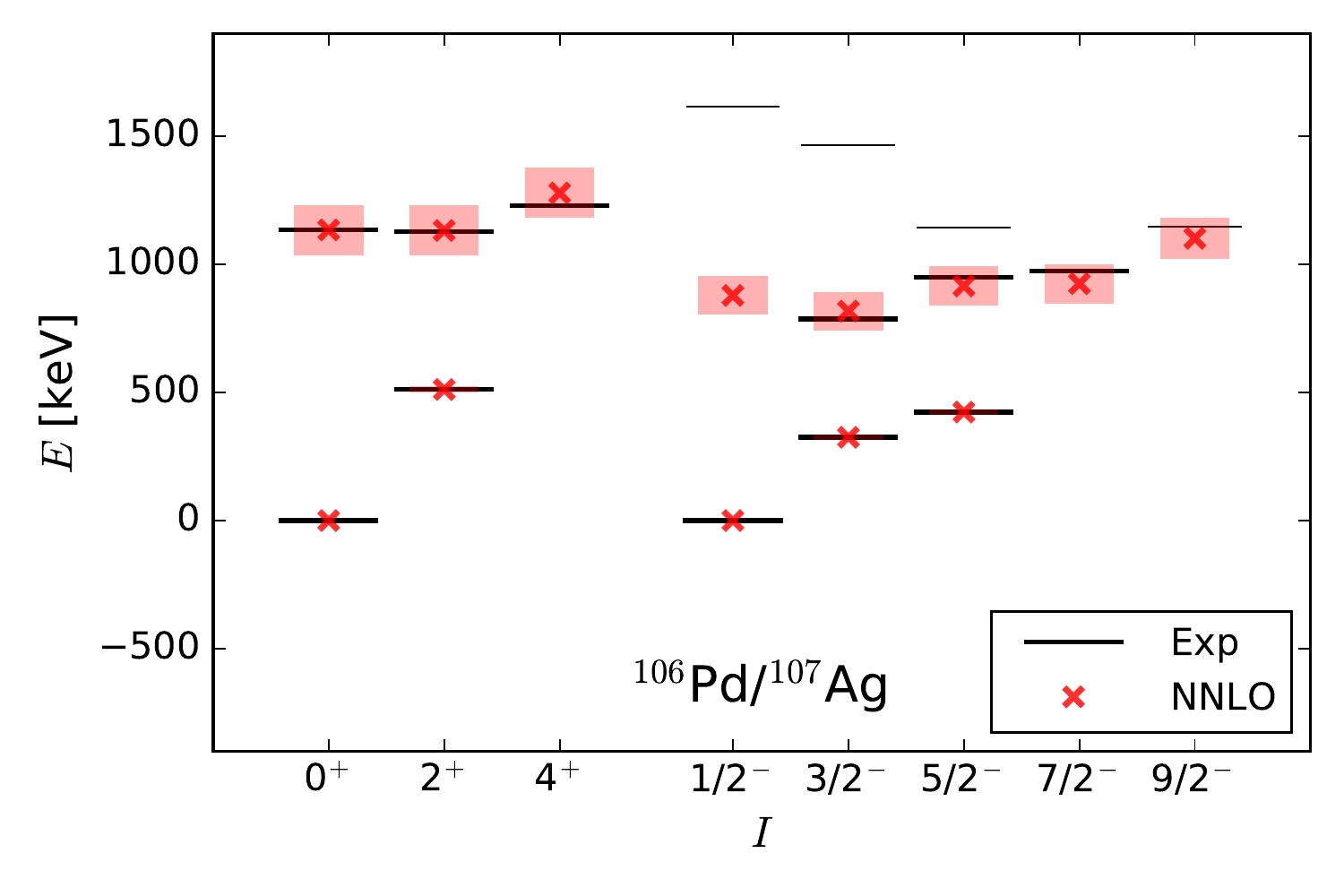}
\includegraphics[width=0.49\textwidth]{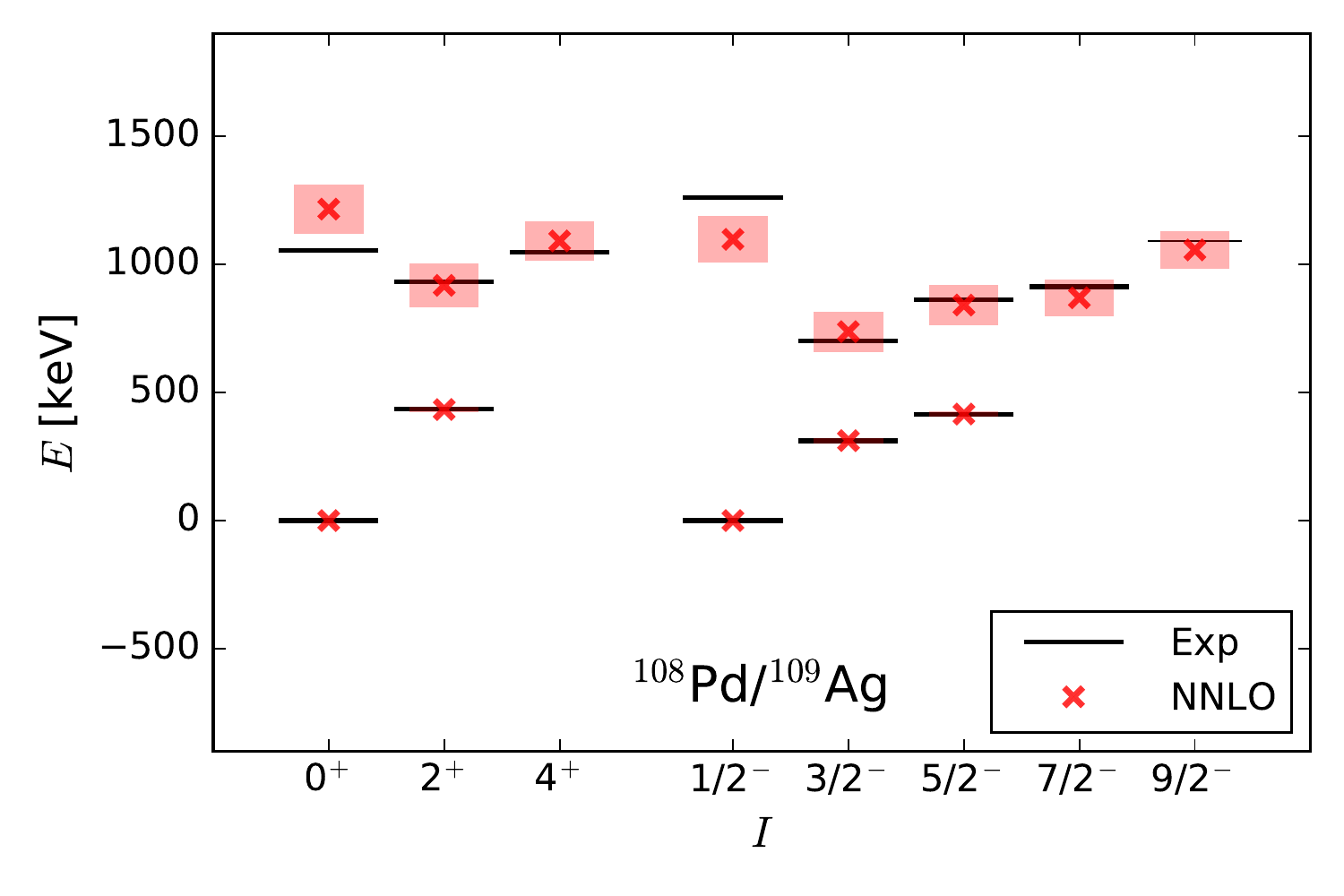}
\includegraphics[width=0.49\textwidth]{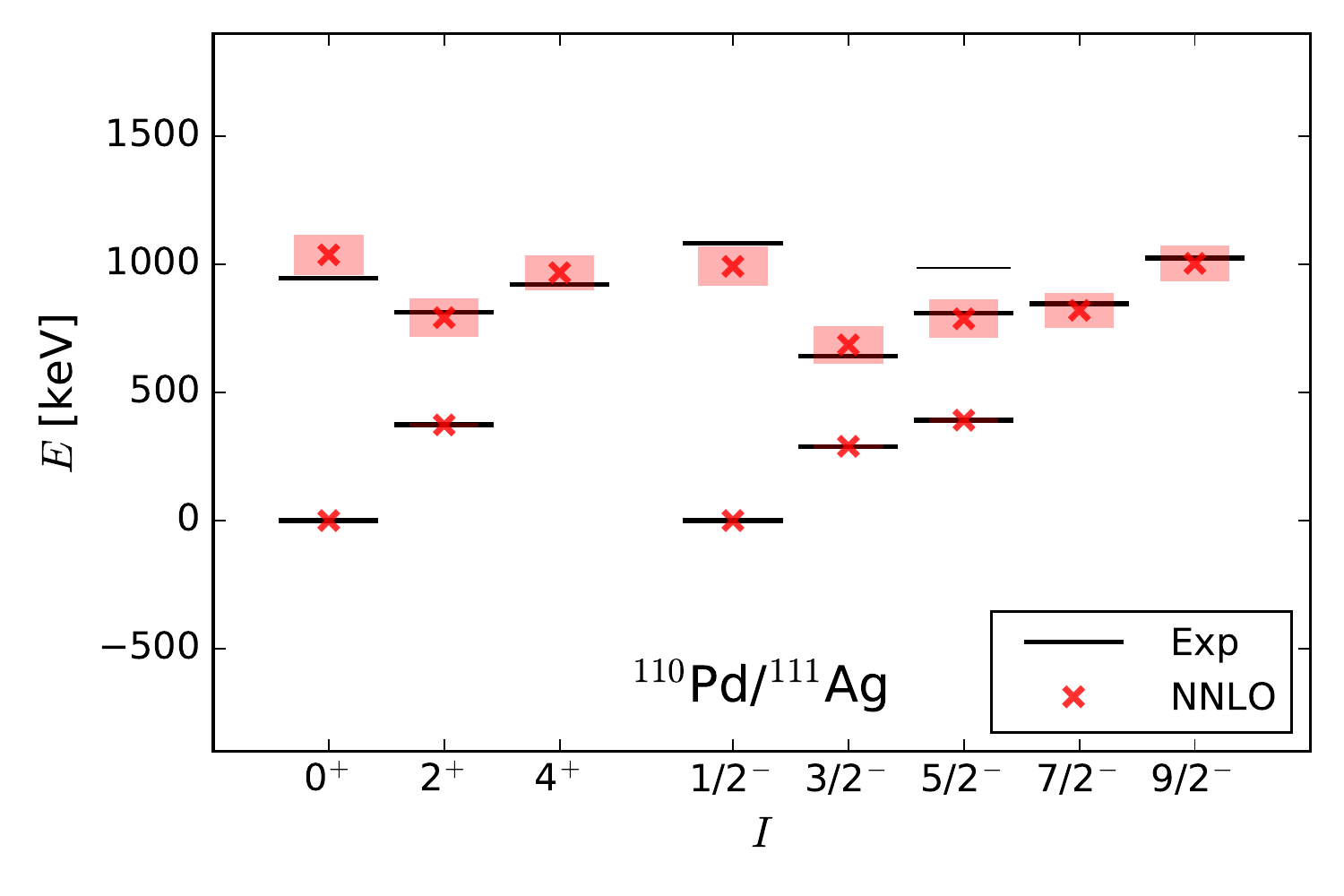}
\caption{(Color online) NNLO energy spectra of Pd/Ag systems. Ag is
  described as a proton in a $j^{\pi}=\sfrac{1}{2}^{-}$ orbital
  coupled to a Pd core. Thick black lines denote states employed to
  fit the LECs while thin black lines denote states with a definitely
  known spin or a single tentative spin-parity assignment. Red crosses
  and shaded areas denote theoretical predictions and uncertainties,
  respectively.}
\label{pdag}
\end{figure*}

\begin{figure}[htb!]
\centering
\includegraphics[width=0.49\textwidth]{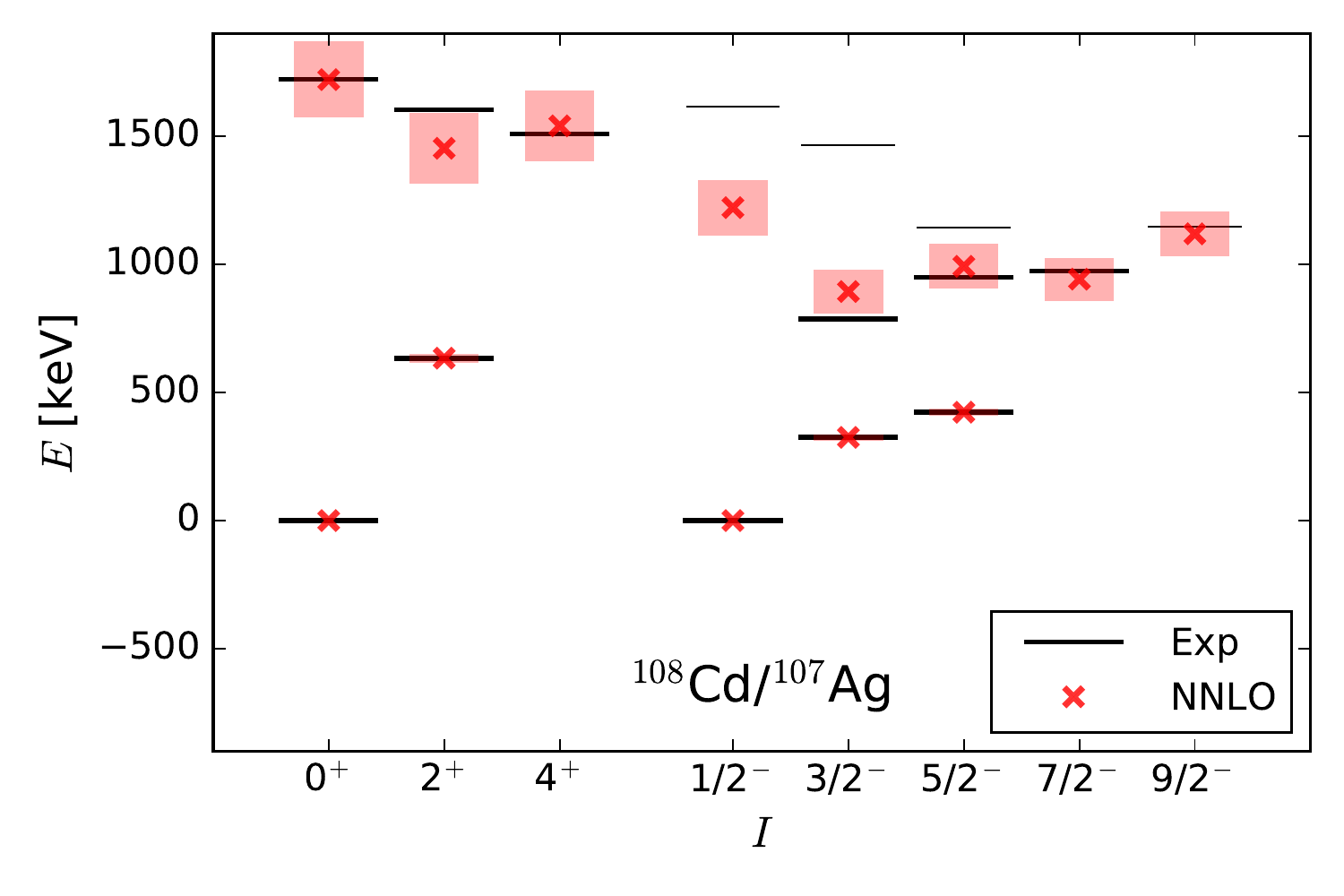}
\includegraphics[width=0.49\textwidth]{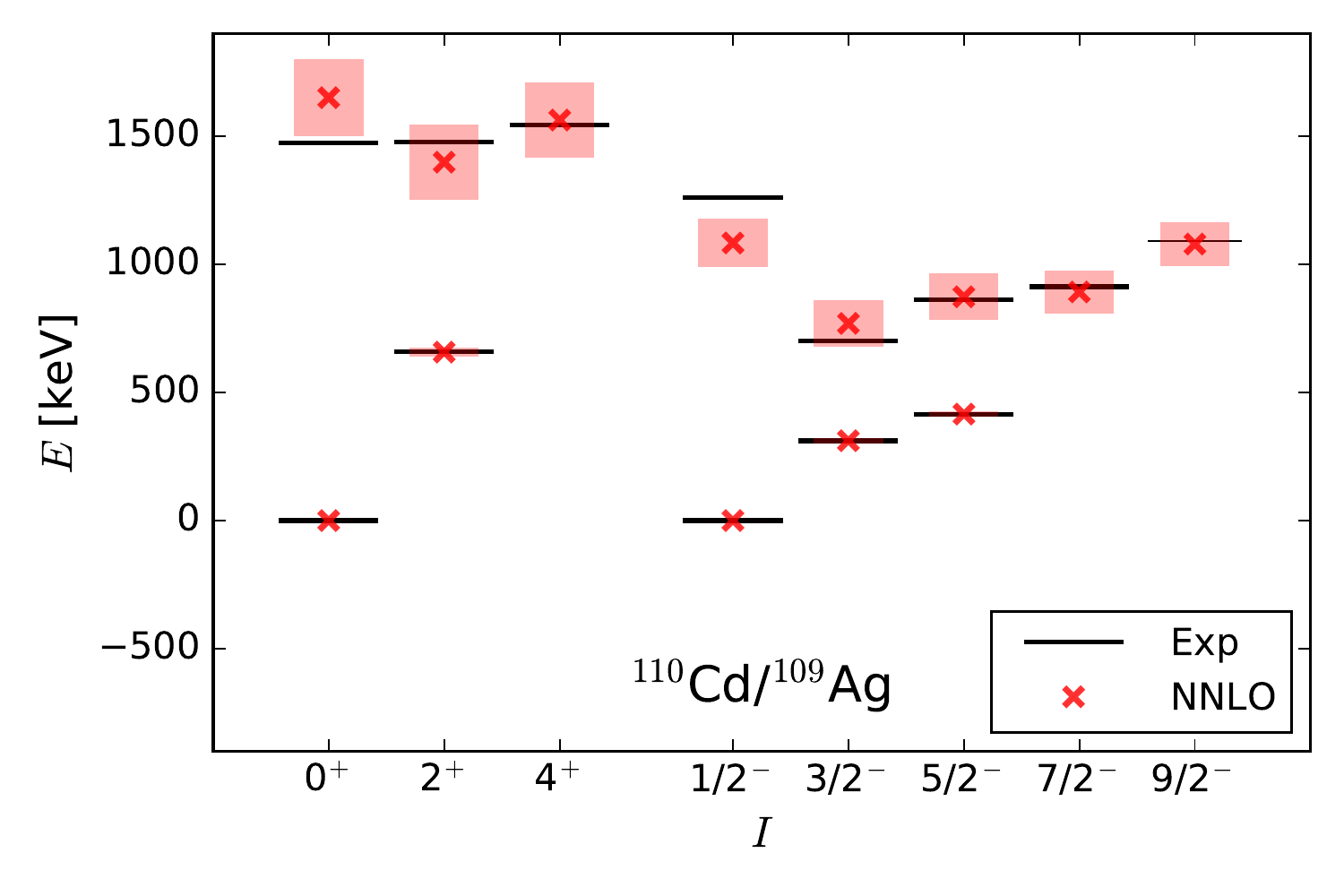}
\includegraphics[width=0.49\textwidth]{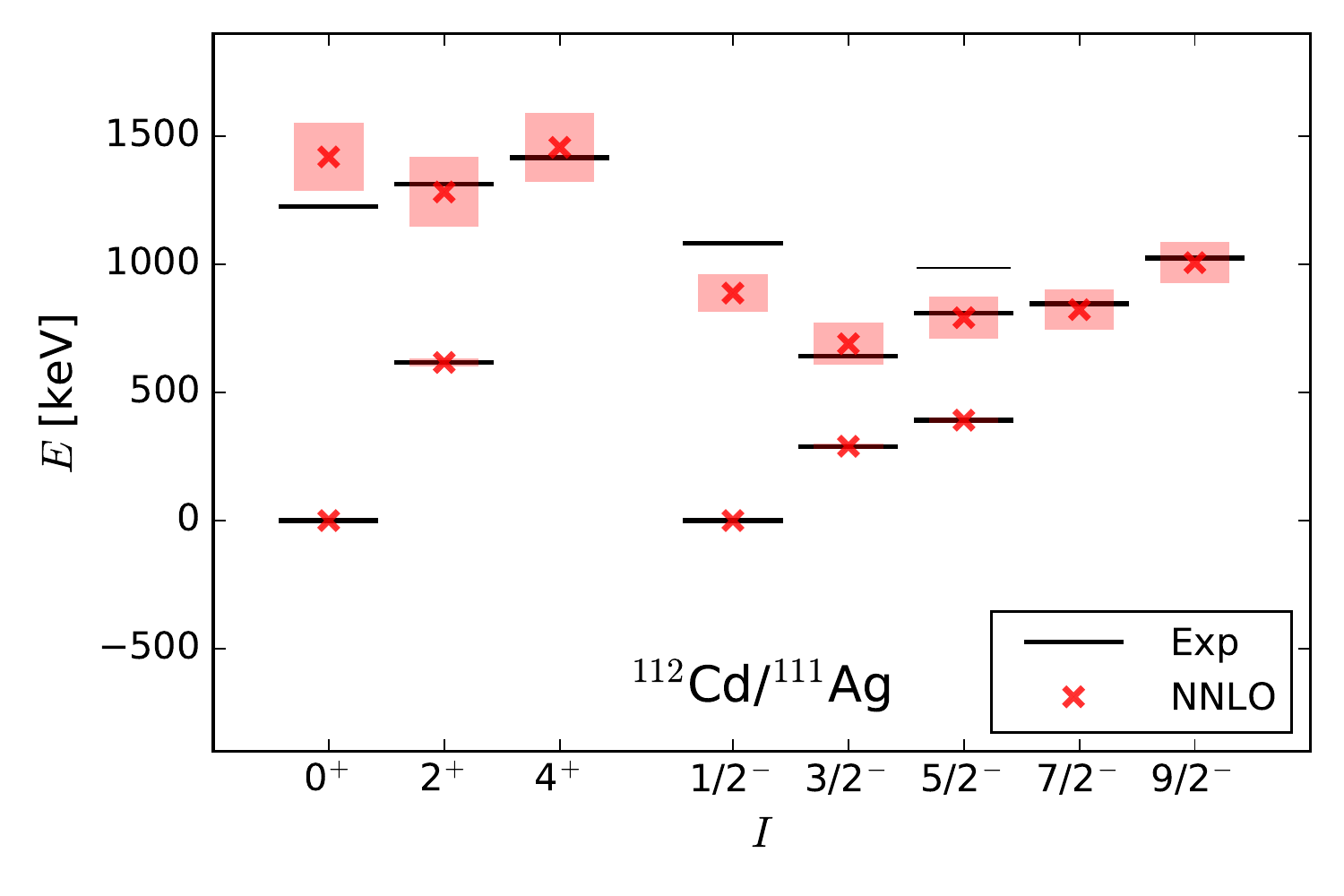}
\caption{(Color online) NNLO energy spectra of Cd/Ag systems. Ag is
  described as a proton hole in a $j^{\pi}=\sfrac{1}{2}^{-}$ orbital
  coupled to a Cd core. Thick black lines denote states employed to
  fit the LECs while thin black lines denote states with a definitely
  known spin or a single tentative spin-parity assignment. Red crosses
  and shaded areas denote theoretical predictions and uncertainties,
  respectively.}
\label{cdag}
\end{figure}

Figures~\ref{rurh}, \ref{pdag}, and~\ref{cdag} show the NNLO energy
spectra of the systems listed in Table~\ref{LOELECs}.
In these figures, even-even and odd-mass states are shown on the left
and right sides, respectively. States employed to fit the LECs are
shown as thick black lines, while additional states (with a definitely
known spin/parity or a single tentative spin/parity assignment) are
shown as thin black lines. Observed levels with more than one
tentative spin/parity assignment are not shown, and we limited
ourselves to negative-parity states in the odd-mass nuclei.  The NNLO
energies~(\ref{NLOE}) for the even-even and odd-mass nuclei are shown
as red crosses. Uncertainties associated to these energies are shown
as red shaded areas. From the power counting, the
next-to-next-to-next-to-leading order (N3LO) corrections to the
energies are expected to scale as $\varepsilon^{3}$, see
Eq.~(\ref{vareps}). The uncertainties associated to the NNLO energies
are quantified using this estimate and the Bayesian method described
in the previous section as
\begin{equation}
\Delta E_{\rm NNLO}(I^{\pi})=\omega_1\delta(I^{\pi}),
\end{equation}
where $\delta$ comes from intervals with a 68\% DOB. Data tables
show several states with tentative spin assignments that would be
consistent with the theoretical results (where no bar is shown).

The comparison of Figs.~\ref{pdag} and \ref{cdag} shows that silver
isotopes can be described either as a proton particle or a proton hole
coupled to palladium or cadmium, respectively.  In the latter case,
theoretical uncertainties are larger than in the former, possibly because
cadmium isotopes have a lower breakdown scale for
vibrations~\cite{coelloperez2015-2}.

\begin{figure}[h!]
\centering
\includegraphics[width=0.49\textwidth]{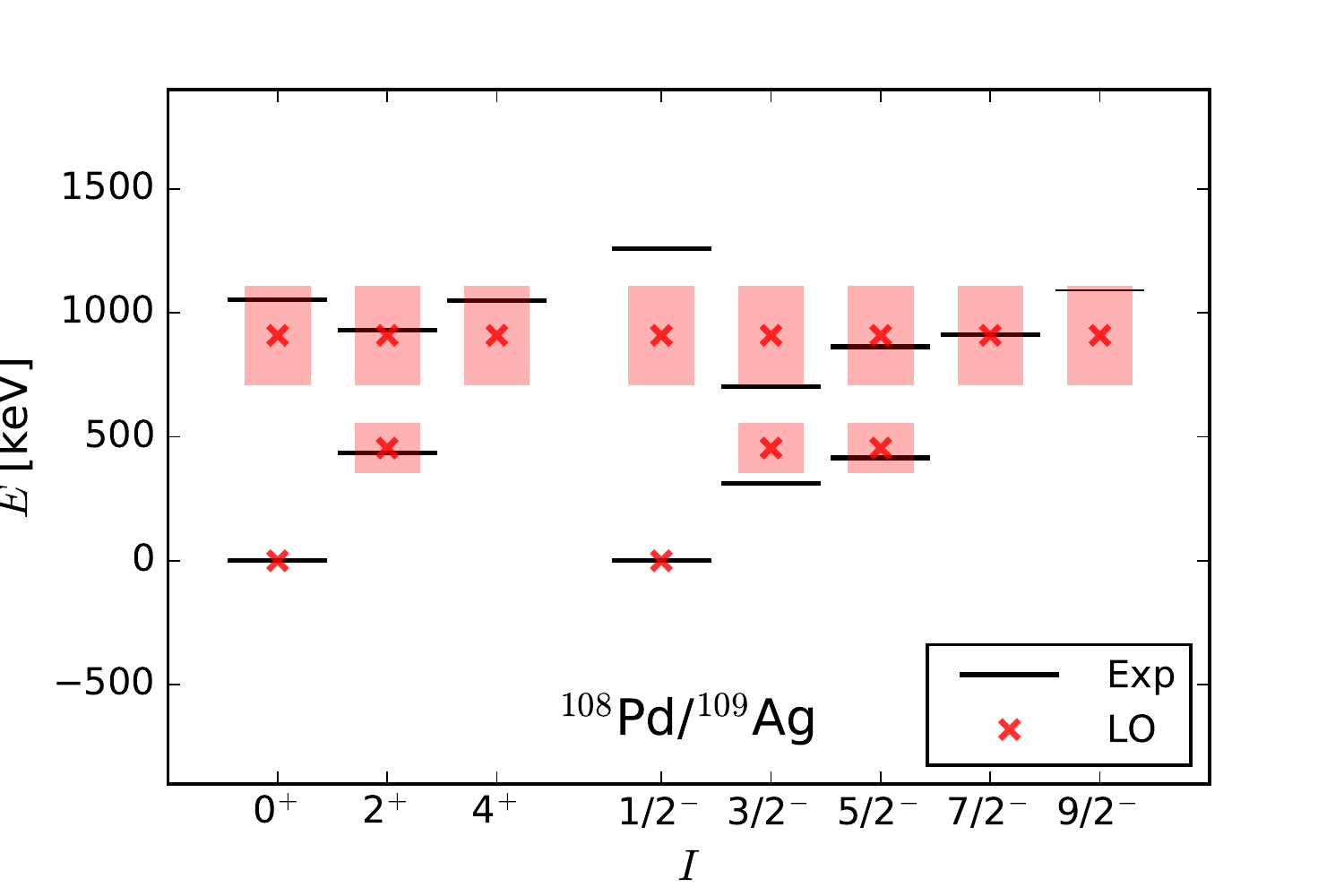}
\includegraphics[width=0.49\textwidth]{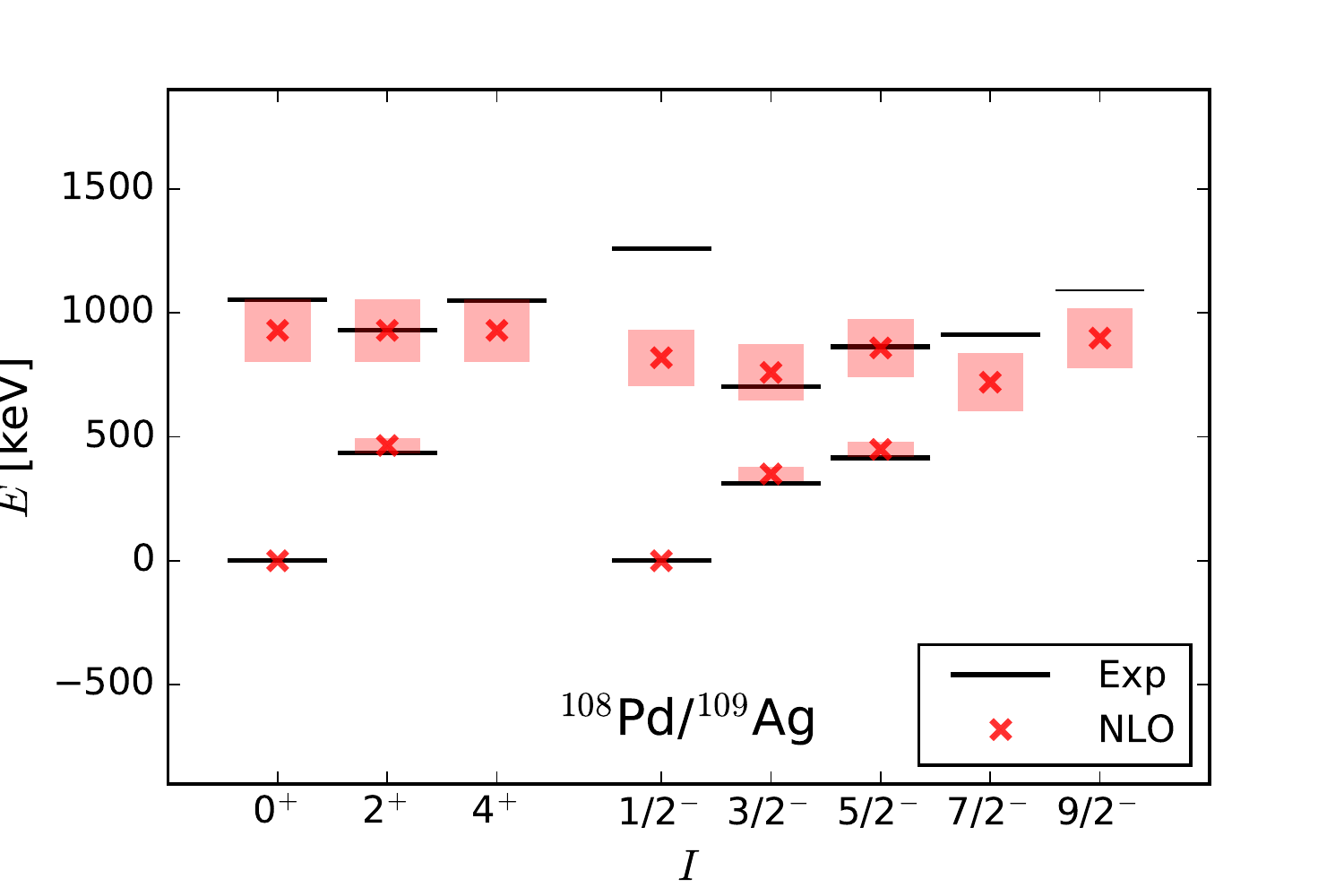}
\includegraphics[width=0.49\textwidth]{108Pd109Ag_NNLOspectrum.pdf}
\caption{(Color online) LO (top), NLO (center) and NNLO (bottom)
  energy spectra of the $^{108}{\rm Pd}$/$^{109}{\rm Ag}$ system. The
  systematic improvement inherent to EFT approaches is evident.}
\label{convergence}
\end{figure}

In order to illustrate the systematic improvement of the EFT we show
the LO, NLO and NNLO energy spectra of the $^{108}{\rm
  Pd}$/$^{109}{\rm Ag}$ system in Figure~\ref{convergence}. The
accuracy (agreement with data) and the precision (decrease of
theoretical uncertainties) increase with increasing order of the
EFT. However, this comes at the cost of reduced predictive power as an
increasing number of LECs need to be adjusted to data.

\section{\textit{E}2 observables}\label{E2}
$E2$ transitions and moments result from the minimal and nonminimal
coupling of the effective degrees of freedom to gauge fields and
electric fields, respectively. Due to Siegert's theorem, coupling to
gauge fields ca also be rewritten as nonminimal couplings. The $E2$
operator is is a spherical tensor of rank two. In this study we are
interested in the reduced $E2$ strengths for transitions between
states differing by none or one phonon, and the static $E2$
moments. The relevant terms of the $E2$ operator for the calculation
of these observables are~\cite{coelloperez2015-2}
\begin{equation}
\hat{Q}_{\mu} = Q_{0}\left(d^{\dagger}_{\mu}+\tilde{d}_{\mu}\right) + Q_{1}\left( d^{\dagger} \otimes \tilde{d} \right)^{(2)}_{\mu}.
\label{E2op}
\end{equation}
Here $Q_{0}$ and $Q_{1}$ are LECs that must be fit to data. From the
power counting one expects $Q_{1}$ to scale as
\begin{equation}
Q_{1}\sim \sqrt{\frac{\omega_1}{\Lambda}}Q_{0}\sim \sqrt{\frac{1}{3}}Q_{0}.
\end{equation}
For the odd-mass nuclei we consider, the $j=\sfrac{1}{2}$ orbital must
couple to boson degrees of freedom to obtain a rank-two tensor. Thus,
we could replace $Q_{0,1}$ in Eq.~(\ref{E2op}) by the linear combination
$q_{0,1} + \tilde{q}_{0,1}\hat{n}$ to include fermion effects. Based
on the power counting [recall the discussion of the
  Hamiltonians~(\ref{Hbf}) and (\ref{ham})], the terms proportional to
$\hat{n}$ are subleading corrections. This agrees with our
expectations: $B(E2)$ strengths associated with collective quadrupole
transitions in even-even nuclei are about tens of Weisskopf units in
size and therefore much larger than single-particle effects.  Here, we
limit ourselves to the leading terms that change and preserve phonon
numbers. In Eq.~(\ref{E2op}) the corresponding operators are
proportional to $Q_0$ and $Q_1$, respectively.

The reduced $E2$ strength or $B(E2)$ value for the transition between
the initial and final states $|i\rangle$ and $|f\rangle$,
respectively, is 
\begin{equation}
B(E2; i\rightarrow f) = \frac{\left|\langle f||\hat{Q}||i\rangle\right|^{2}}{2I_{i}+1} .
\label{BE2}
\end{equation}
Here, 
\begin{equation}
\langle f||\hat{O}||i\rangle = \frac{\sqrt{2I_{f}+1}}{C_{I_{i}M_{i}\lambda M_{f}-M_{i}}^{I_{f}M_{f}}} \langle f|\hat{O}_{M_{f}-M_{i}}|i\rangle
\end{equation}
is the reduced matrix element of an spherical operator $\hat{O}$ of rank $\lambda$.
The static $E2$ moment of the state $I_{i}$ is defined as~\cite{rowe2010}
\begin{equation}
Q(I_{i})= \sqrt{\frac{16\pi}{5}} \frac{C_{II20}^{II}}{\sqrt{2I+1}}\langle I_{i} || \hat{Q} || I_{i} \rangle.
\label{E2momdef}
\end{equation}
This definition is consistent when comparing the diagonal reduced matrix elements of the $E2$ operator in $^{106}{\rm Pd}$ and $^{108}{\rm Pd}$ reported in Ref.~\cite{svensson1995} and the static $E2$ moments for the same nuclei reported in Ref.~\cite{stone2014}.

\subsection{Phonon-annihilating transition strengths}
The power counting establishes the transitions between states
differing by one phonon as the strongest $E2$ observables. In what
follows we discuss transitions in which one phonon is annihilated. The term
proportional to $Q_{0}$ in the $E2$ operator~(\ref{E2op}) couples states
that differ by one phonon; thus, the $E2$ transition strengths for
one-phonon decays are governed by this LEC. The reduced matrix
elements required for their calculation are
\begin{widetext}
\begin{equation}
\begin{aligned}
\langle I'; N-1; 0 || \hat{Q} || I; N; 0 \rangle &= Q_{0} \sqrt{N}\Pi_I \quad\mbox{for}~N=1,2 \\ \\
\langle I'; N-1; \tfrac{1}{2} || \hat{Q} || I; NJ; \tfrac{1}{2} \rangle
&= \left\{\begin{array}{l c} 
Q_{0} \Pi_I & \mbox{for}~N=1 \\ \\
Q_{0} (-1)^{I'+\tfrac{1}{2}} \sqrt{2}\Pi_{I'JI}
\left\{ \begin{array}{c c c} 2 & 2 & J \\ \tfrac{1}{2} & I & I' \end{array} \right\} &
\mbox{for}~N=2 \end{array} \right.
\end{aligned}
\label{Q0matele}
\end{equation}
\end{widetext}
Here we used the shorthand
\begin{equation}
\Pi_{ab\ldots c} \equiv \sqrt{(2a+1)(2b+1)\ldots(2c+1)}.
  \end{equation}

Table~\ref{LOredE2} lists the reduced matrix elements for the
transitions of interest resulting from Eq.~(\ref{Q0matele}) in terms
of the LEC $Q_{0}$.
\begin{table}
\caption{Reduced matrix elements relevant for phonon-annihilating
  transitions in units of $Q_{0}$.}  \centering
\begin{tabular}{ccc}
\hline\hline
System & $I_{i} \rightarrow I_{f}$ & $\langle f||\hat{Q}||i\rangle$ \\
\hline
even-even & $2_{1}\rightarrow 0_{1}$ & $\sqrt{5}$ \\
 & $0_{2}\rightarrow 2_{1}$ & $\sqrt{2}$ \\
 & $2_{2}\rightarrow 2_{1}$ & $\sqrt{10}$ \\
 & $4_{1}\rightarrow 2_{1}$ & $\sqrt{18}$ \\
\hline
odd-mass & $\frac{3}{2}_{1} \rightarrow \frac{1}{2}_{1}$ & 2 \\
 & $\frac{5}{2}_{1} \rightarrow \frac{1}{2}_{1}$ & $\sqrt{6}$ \\
 & $\frac{1}{2}_{2} \rightarrow \frac{3}{2}_{1}$ & $-\sqrt{\frac{8}{5}}$ \\
 & $\frac{1}{2}_{2} \rightarrow \frac{5}{2}_{1}$ & $\sqrt{\frac{12}{5}}$ \\
 & $\frac{3}{2}_{2} \rightarrow \frac{3}{2}_{1}$ & $\sqrt{\frac{28}{5}}$ \\
 & $\frac{3}{2}_{2} \rightarrow \frac{5}{2}_{1}$ & $\sqrt{\frac{12}{5}}$ \\
 & $\frac{5}{2}_{2} \rightarrow \frac{3}{2}_{1}$ & $-\sqrt{\frac{12}{5}}$ \\
 & $\frac{5}{2}_{2} \rightarrow \frac{5}{2}_{1}$ & $\sqrt{\frac{48}{5}}$ \\
 & $\frac{7}{2}_{1} \rightarrow \frac{3}{2}_{1}$ & $\sqrt{\frac{72}{5}}$ \\
 & $\frac{7}{2}_{1} \rightarrow \frac{5}{2}_{1}$ & $\sqrt{\frac{8}{5}}$ \\
 & $\frac{9}{2}_{1} \rightarrow \frac{5}{2}_{1}$ & $\sqrt{20}$ \\
\hline\hline
\end{tabular}
\label{LOredE2}
\end{table}
NLO corrections to these matrix elements are expected to scale as
$\varepsilon$. As a cautionary note we remark that identical units have to
be employed when fitting $Q_{0}$ to experimental data, and
recall that Weisskopf units depend on the number of nucleons $A$ for
$E2$ transitions.

Tables~\ref{102-103LOBE2}, ~\ref{106-107LOBE2}, ~\ref{108-109LOBE2}
and~\ref{110-109LOBE2} show LO results for phonon-annihilating $E2$
transition strengths in the $^{102}{\rm Ru}$/$^{103}{\rm Rh}$,
$^{106}{\rm Pd}$/$^{107}{\rm Ag}$, $^{108}{\rm Pd}$/$^{109}{\rm Ag}$,
and $^{110}{\rm Cd}$/$^{109}{\rm Ag}$, respectively.  The
uncertainties in these tables are quantified as $Q_{0}^{2}\delta$,
and $\delta$ comes from 68\% DOB intervals.  For the other systems
studied in this work, data on $E2$ transition strengths are
insufficient to conduct a similar analysis.  It would be valuable to
measure $E2$ transition strengths in those systems in order to further
test the EFT.

\begin{table}
\caption{Reduced transition probabilities for phonon-annihilating $E2$
  transitions in the $^{102}{\rm Ru}$/$^{103}{\rm Rh}$ system in
  Weisskopf units. The uncertainty was quantified from 68\% DOB
  intervals.}  \centering
\begin{tabular}{ccd{5.0}d{5.0}}
\hline\hline
Nucleus & $I_{i}^{\pi}\rightarrow I_{f}^{\pi}$ & \multicolumn{1}{c}
{$B(E2)_{\rm exp}$} & \multicolumn{1}{c}{$B(E2)_{\rm EFT}$} \\
\hline
$^{102}{\rm Ru}$ & $2^{+}_{1}\rightarrow 0^{+}_{1}$ & 45(1) & 27(9) \\
 & $0^{+}_{2}\rightarrow 2^{+}_{1}$ & 35(6) & 55(18) \\
 & $2^{+}_{2}\rightarrow 2^{+}_{1}$ & 32(5) & 55(18) \\
 & $4^{+}_{1}\rightarrow 2^{+}_{1}$ & 66(11) & 55(18) \\
\hline
$^{103}{\rm Rh}$ & $\frac{3}{2}^{-}_{1}\rightarrow\frac{1}{2}^{-}_{1}$ &
36(4) & 27(9) \\
 & $\frac{5}{2}^{-}_{1}\rightarrow\frac{1}{2}^{-}_{1}$ & 44(3) & 27(9) \\
 & $\frac{1}{2}^{-}_{2}\rightarrow\frac{3}{2}^{-}_{1}$ & & 22(18) \\
 & $\frac{1}{2}^{-}_{2}\rightarrow\frac{5}{2}^{-}_{1}$ & 486(90) & 32(18)
 \\
 & $\frac{3}{2}^{-}_{2}\rightarrow\frac{3}{2}^{-}_{1}$ & & 38(18) \\
 & $\frac{3}{2}^{-}_{2}\rightarrow\frac{5}{2}^{-}_{1}$ & & 16(18) \\
 & $\frac{5}{2}^{-}_{2}\rightarrow\frac{3}{2}^{-}_{1}$ & 3(1) & 11(18) \\
 & $\frac{5}{2}^{-}_{2}\rightarrow\frac{5}{2}^{-}_{1}$ & 4(1) & 43(18) \\
 & $\frac{7}{2}^{-}_{1}\rightarrow\frac{3}{2}^{-}_{1}$ & 34(11) & 48(18) \\
 & $\frac{7}{2}^{-}_{1}\rightarrow\frac{5}{2}^{-}_{1}$ & & 5(18) \\
 & $\frac{9}{2}^{-}_{1}\rightarrow\frac{5}{2}^{-}_{1}$ & 46(7) & 54(18) \\
\hline\hline
\end{tabular}
\label{102-103LOBE2}
\end{table}

\begin{table}
\caption{Reduced transition probabilities for phonon-annihilating $E2$
  transitions in the $^{106}{\rm Pd}$/$^{107}{\rm Ag}$ system in
  Weisskopf units. The uncertainty was quantified from 68\% DOB
  intervals.}  \centering
\begin{tabular}{ccd{5.0}d{5.0}}
\hline\hline
Nucleus & $I_{i}^{\pi}\rightarrow I_{f}^{\pi}$ & \multicolumn{1}{c}{$B(E2)_{\rm exp}$} & \multicolumn{1}{c}{$B(E2)_{\rm EFT}$} \\
\hline
$^{106}{\rm Pd}$ & $2^{+}_{1}\rightarrow 0^{+}_{1}$ & 44(1) & 35(12) \\
 & $0^{+}_{2}\rightarrow 2^{+}_{1}$ & 35(8) & 69(23) \\
 & $2^{+}_{2}\rightarrow 2^{+}_{1}$ & 44(4) & 69(23) \\
 & $4^{+}_{1}\rightarrow 2^{+}_{1}$ & 76(11) & 69(23) \\
\hline
$^{107}{\rm Ag}$ & $\frac{3}{2}^{-}_{1}\rightarrow\frac{1}{2}^{-}_{1}$ & 42(4) & 34(11) \\
 & $\frac{5}{2}^{-}_{1}\rightarrow\frac{1}{2}^{-}_{1}$ & 43(3) & 34(11) \\
 & $\frac{1}{2}^{-}_{2}\rightarrow\frac{3}{2}^{-}_{1}$ & & 27(23) \\
 & $\frac{1}{2}^{-}_{2}\rightarrow\frac{5}{2}^{-}_{1}$ & & 41(23) \\
 & $\frac{3}{2}^{-}_{2}\rightarrow\frac{3}{2}^{-}_{1}$ & & 48(23) \\
 & $\frac{3}{2}^{-}_{2}\rightarrow\frac{5}{2}^{-}_{1}$ & & 20(23) \\
 & $\frac{5}{2}^{-}_{2}\rightarrow\frac{3}{2}^{-}_{1}$ & & 14(23) \\
 & $\frac{5}{2}^{-}_{2}\rightarrow\frac{5}{2}^{-}_{1}$ & & 55(23) \\
 & $\frac{7}{2}^{-}_{1}\rightarrow\frac{3}{2}^{-}_{1}$ & & 62(23) \\
 & $\frac{7}{2}^{-}_{1}\rightarrow\frac{5}{2}^{-}_{1}$ & & 7(23) \\
 & $\frac{9}{2}^{-}_{1}\rightarrow\frac{5}{2}^{-}_{1}$ & & 68(23) \\
\hline\hline
\end{tabular}
\label{106-107LOBE2}
\end{table}

\begin{table}
\caption{Reduced transition probabilities for phonon-annihilating $E2$
  transitions in the $^{108}{\rm Pd}$/$^{109}{\rm Ag}$ system in
  Weisskopf units. The uncertainty was quantified from 68\% DOB
  intervals.}  \centering
\begin{tabular}{ccd{5.0}d{5.0}}
\hline\hline
Nucleus & $I_{i}^{\pi}\rightarrow I_{f}^{\pi}$ & \multicolumn{1}{c}{$B(E2)_{\rm exp}$} & \multicolumn{1}{c}{$B(E2)_{\rm EFT}$} \\
\hline
$^{108}{\rm Pd}$ & $2^{+}_{1}\rightarrow 0^{+}_{1}$ & 49(1) & 34(11) \\
 & $0^{+}_{2}\rightarrow 2^{+}_{1}$ & 52(5) & 69(23) \\
 & $2^{+}_{2}\rightarrow 2^{+}_{1}$ & 71(5) & 69(23) \\
 & $4^{+}_{1}\rightarrow 2^{+}_{1}$ & 73(8) & 69(23) \\
\hline
$^{109}{\rm Ag}$ & $\frac{3}{2}^{-}_{1}\rightarrow\frac{1}{2}^{-}_{1}$ & 40(40) & 34(11) \\
 & $\frac{5}{2}^{-}_{1}\rightarrow\frac{1}{2}^{-}_{1}$ & 41(6) & 34(11) \\
 & $\frac{1}{2}^{-}_{2}\rightarrow\frac{3}{2}^{-}_{1}$ & & 27(23) \\
 & $\frac{1}{2}^{-}_{2}\rightarrow\frac{5}{2}^{-}_{1}$ & & 41(23) \\
 & $\frac{3}{2}^{-}_{2}\rightarrow\frac{3}{2}^{-}_{1}$ & 49(24) & 47(23) \\
 & $\frac{3}{2}^{-}_{2}\rightarrow\frac{5}{2}^{-}_{1}$ & & 20(23) \\
 & $\frac{5}{2}^{-}_{2}\rightarrow\frac{3}{2}^{-}_{1}$ & 8(4) & 14(23) \\
 & $\frac{5}{2}^{-}_{2}\rightarrow\frac{5}{2}^{-}_{1}$ & 10(7) & 54(23) \\
 & $\frac{7}{2}^{-}_{1}\rightarrow\frac{3}{2}^{-}_{1}$ & & 61(23) \\
 & $\frac{7}{2}^{-}_{1}\rightarrow\frac{5}{2}^{-}_{1}$ & & 7(23) \\
 & $\frac{9}{2}^{-}_{1}\rightarrow\frac{5}{2}^{-}_{1}$ & & 68(23) \\
\hline\hline
\end{tabular}
\label{108-109LOBE2}
\end{table}
\begin{table}
\caption{Reduced transition probabilities for phonon-annihilating $E2$
  transitions in the $^{110}{\rm Cd}$/$^{109}{\rm Ag}$ system in
  Weisskopf units. The uncertainty was quantified from 68\% DOB
  intervals.}  \centering
\begin{tabular}{ccd{5.0}d{5.0}}
\hline\hline
Nucleus & $I_{i}^{\pi}\rightarrow I_{f}^{\pi}$ & \multicolumn{1}{c}{$B(E2)_{\rm exp}$} & \multicolumn{1}{c}{$B(E2)_{\rm EFT}$} \\
\hline
$^{110}{\rm Cd}$ & $2^{+}_{1}\rightarrow 0^{+}_{1}$ & 27(1) & 23(8) \\
 & $0^{+}_{2}\rightarrow 2^{+}_{1}$ & & 46(15) \\
 & $2^{+}_{2}\rightarrow 2^{+}_{1}$ & 30(5) & 46(15) \\
 & $4^{+}_{1}\rightarrow 2^{+}_{1}$ & 42(9) & 46(15) \\
\hline
$^{109}{\rm Ag}$ & $\frac{3}{2}^{-}_{1}\rightarrow\frac{1}{2}^{-}_{1}$ & 40(40) & 23(8) \\
 & $\frac{5}{2}^{-}_{1}\rightarrow\frac{1}{2}^{-}_{1}$ & 41(6) & 23(8) \\
 & $\frac{1}{2}^{-}_{2}\rightarrow\frac{3}{2}^{-}_{1}$ & & 19(16) \\
 & $\frac{1}{2}^{-}_{2}\rightarrow\frac{5}{2}^{-}_{1}$ & & 28(16) \\
 & $\frac{3}{2}^{-}_{2}\rightarrow\frac{3}{2}^{-}_{1}$ & 49(24) & 33(16) \\
 & $\frac{3}{2}^{-}_{2}\rightarrow\frac{5}{2}^{-}_{1}$ & & 14(16) \\
 & $\frac{5}{2}^{-}_{2}\rightarrow\frac{3}{2}^{-}_{1}$ & 8(4) & 9(16) \\
 & $\frac{5}{2}^{-}_{2}\rightarrow\frac{5}{2}^{-}_{1}$ & 10(7) & 37(16) \\
 & $\frac{7}{2}^{-}_{1}\rightarrow\frac{3}{2}^{-}_{1}$ & & 42(16) \\
 & $\frac{7}{2}^{-}_{1}\rightarrow\frac{5}{2}^{-}_{1}$ & & 5(16) \\
 & $\frac{9}{2}^{-}_{1}\rightarrow\frac{5}{2}^{-}_{1}$ & & 47(16) \\
\hline\hline
\end{tabular}
\label{110-109LOBE2}
\end{table}

Most of the available data on $E2$ transition strengths were employed
to fit the single LEC $Q_{0}$. The only exception was the
$(\sfrac{1}{2})^{1}_{2}\rightarrow (\sfrac{5}{2})^{-}_{1}$ transition strength
in $^{103}{\rm Rh}$, which was excluded due to its unexpectedly large
value. The values of $Q_{0}$ for the $^{102}{\rm Ru}$/$^{103}{\rm Rh}$,
$^{106}{\rm Pd}$/$^{107}{\rm Ag}$,
$^{108}{\rm Pd}$/$^{109}{\rm Ag}$ and
$^{110}{\rm Cd}$/$^{109}{\rm Ag}$ systems are $0.28$, $0.32$,
$0.32$ and $0.27$~eb, respectively. Note that the transition strengths in
$^{109}{\rm Ag}$ can be described employing either$^{108}{\rm Pd}$
or $^{110}{\rm Cd}$ as a core. Both descriptions agree with each
other within theoretical uncertainties.

\subsection{Static moments and phonon-conserving transition strengths}

The term proportional to $Q_1$ in the $E2$ operator~(\ref{E2op})
couples states with the same number of phonons. Thus, $Q_{1}$
enters in the LO calculation of static $E2$ moments. The reduced
matrix elements associated to these observables are
\begin{widetext}
\begin{equation}
\begin{aligned}
\langle I'; N; 0 || \hat{Q} || I; N; 0 \rangle &= \left\{ \begin{array}{l c}
0 &\mbox{for}~N=0 \\ \\
Q_{1} \Pi_I &\mbox{for}~N=1 \\ \\
2Q_{1} \sqrt{5}\Pi_{I'I} \left\{ \begin{array}{c c c} 2 & 2 & 2 \\
2 & I & I' \end{array} \right\} &\mbox{for}~N=2 \end{array} \right. \\ \\
\langle I'; NJ'; \tfrac{1}{2} || \hat{Q} || I; NJ; \tfrac{1}{2} \rangle &= \left\{ \begin{array}{l c}
0 &\mbox{for} N=0 \\ \\
Q_{1}(-1)^{I+\tfrac{1}{2}}\sqrt{5}\Pi_{I'I} \left\{ \begin{array}{c c c}
2 & 2 & 2 \\ \tfrac{1}{2} & I & I' \end{array} \right\} &\mbox{for}~N=1 \\ \\
2Q_{1}(-1)^{I+\tfrac{1}{2}}\sqrt{5}\Pi_{I'J'IJ}
\left\{\begin{array}{c c c} 2 & 2 & 2 \\ 2 & J' & J \end{array} \right\}
\left\{\begin{array}{c c c} 2 & J' & J \\ \tfrac{1}{2} & I & I' \end{array} \right\} &\mbox{for}~N=2
\end{array} \right.
\end{aligned}
\label{Q1matele}
\end{equation}
\end{widetext}

The static $E2$ moments and reduced matrix elements required to
calculate the $E2$ strengths for transitions between two-phonon states
at LO are given in units of $Q_{1}$ in Tables~\ref{E2mom}
and~\ref{NLOredE2}, respectively.
\begin{table}
\caption{Static $E2$ moments of states up to the two-phonon
  level in units of $Q_{1}$. The subindex $i$ indicates the position
  of the excited state.}  \centering
\begin{tabular}{ccc}
\hline\hline
System & $I_{i}$ & $Q(I_{i})$ \\
\hline
even-even & $2_{1}$ & $8\sqrt{\frac{2\pi}{35}}$ \\
 & $2_{2}$ & $-\frac{24}{7}\sqrt{\frac{2\pi}{35}}$ \\
 & $4_{1}$ & $16\sqrt{\frac{2\pi}{35}}$ \\
\hline
odd-mass & $\frac{3}{2}_{1}$ & $\frac{28}{5}\sqrt{\frac{2\pi}{35}}$ \\
 & $\frac{5}{2}_{1}$ & $8\sqrt{\frac{2\pi}{35}}$ \\
 & $\frac{3}{2}_{2}$ & $-\frac{12}{5}\sqrt{\frac{2\pi}{35}}$ \\
 & $\frac{5}{2}_{2}$ & $-\frac{24}{7}\sqrt{\frac{2\pi}{35}}$ \\
 & $\frac{7}{2}_{1}$ & $\frac{44}{3}\sqrt{\frac{2\pi}{35}}$ \\
 & $\frac{9}{2}_{1}$ & $16\sqrt{\frac{2\pi}{35}}$ \\
\hline\hline
\end{tabular}
\label{E2mom}
\end{table}
\begin{table}
\caption{Reduced matrix elements relevant for phonon-conserving $E2$
  transitions in units of $Q_{1}$.}  \centering
\begin{tabular}{ccc}
\hline\hline
System & $I_{i} \rightarrow I_{f}$ & $\langle f||\hat{Q}||i\rangle$ \\
\hline
even-even & $2_{2}\rightarrow 0_{2}$ & $4$ \\
 & $4_{1}\rightarrow 2_{2}$ & $\frac{24}{7}$ \\
\hline
odd-mass & $\frac{5}{2}_{1} \rightarrow \frac{3}{2}_{1}$ & $-\sqrt{\frac{24}{5}}$ \\
 & $\frac{3}{2}_{2} \rightarrow \frac{1}{2}_{2}$ & $\sqrt{\frac{64}{5}}$ \\
 & $\frac{5}{2}_{2} \rightarrow \frac{1}{2}_{2}$ & $\sqrt{\frac{96}{5}}$ \\
 & $\frac{5}{2}_{2} \rightarrow \frac{3}{2}_{2}$ & $\sqrt{\frac{216}{245}}$ \\
 & $\frac{7}{2}_{1} \rightarrow \frac{3}{2}_{2}$ & $\sqrt{\frac{2304}{245}}$ \\
 & $\frac{7}{2}_{1} \rightarrow \frac{5}{2}_{2}$ & $\sqrt{\frac{256}{245}}$ \\
 & $\frac{9}{2}_{1} \rightarrow \frac{5}{2}_{2}$ & $\sqrt{\frac{640}{49}}$ \\
 & $\frac{9}{2}_{1} \rightarrow \frac{7}{2}_{1}$ & $-\sqrt{\frac{880}{147}}$ \\
\hline\hline
\end{tabular}
\label{NLOredE2}
\end{table}
NLO corrections to these quantities are expected to scale as $\varepsilon$.

Our results for static $E2$ moments in the
$^{102}{\rm Ru}$/$^{103}{\rm Rh}$,
$^{106}{\rm Pd}$/$^{107}{\rm Ag}$ and
$^{108}{\rm Pd}$/$^{109}{\rm Ag}$ systems are listed in
Table~\ref{E2res}, where the theoretical uncertainty for the state $I^{\pi}$
was quantified as $\sqrt{16\pi/5(2I+1)}C_{II20}^{II}Q_{0}\delta$ [in
agreement with the definition given in Eq.~(\ref{E2momdef})], with
$\delta$ from 68\% DOB intervals.
\begin{table}
\caption{Static $E2$ moments in some systems in eb. The uncertainty 
was quantified from 68\% DOB intervals.}
\centering
\begin{tabular}{ccd{2.5}d{2.5}}
\hline\hline
Nucleus & $I_{i}^{\pi}$ & \multicolumn{1}{c}{$Q_{\rm exp}$} &
\multicolumn{1}{c}{$Q_{\rm EFT}$} \\
\hline
$^{102}{\rm Ru}$ & $2_{1}^{+}$ & -0.63 (3) & -0.41 (6) \\
 & $2_{2}^{+}$ & & 0.18(18) \\
 & $4_{1}^{+}$ & & -0.82(14) \\
$^{103}{\rm Rh}$ & $\frac{3}{2}_{1}^{-}$ & -0.3(2) & -0.29(7) \\
 & $\frac{5}{2}_{1}^{-}$ & -0.4(2) & -0.41(6) \\
\hline
$^{106}{\rm Pd}$ & $2_{1}^{+}$ & -0.54 (4) & -0.50 (7) \\
 & $2_{2}^{+}$ & 0.39(6) & 0.21(20) \\
 & $4_{1}^{+}$ & -0.79(11) & -1.00(17) \\
$^{107}{\rm Ag}$ & $\frac{3}{2}_{1}^{-}$ & & -0.35 (8) \\
 & $\frac{5}{2}_{1}^{-}$ & & -0.50(7) \\
\hline$^{108}{\rm Pd}$ & $2_{1}^{+}$ & -0.56(3) & -0.57(7) \\
 & $2_{2}^{+}$ & 0.73(9) & 0.24(20) \\
 & $4_{1}^{+}$ & -0.78(11) & -1.14(17) \\
$^{109}{\rm Ag}$ & $\frac{3}{2}_{1}^{-}$ & -0.7(3) & -0.40(8) \\
 & $\frac{5}{2}_{1}^{-}$ & -0.3(3) & -0.57(6) \\
\hline$^{110}{\rm Cd}$ & $2_{1}^{+}$ & -0.39(3) & -0.57(7) \\
 & $2_{2}^{+}$ & & 0.24(17) \\
 & $4_{1}^{+}$ & & -1.12(14) \\
$^{109}{\rm Ag}$ & $\frac{3}{2}_{1}^{-}$ & -0.7(3) & -0.39(6) \\
 & $\frac{5}{2}_{1}^{-}$ & -0.3(3) & -0.56(6) \\
\hline\hline
\end{tabular}
\label{E2res}
\end{table}
All available data from Refs.~\cite{svensson1995, stone2014} were used
to fit the LEC $Q_{1}$ through weighted averages. Note that for the
studied systems the values for $|Q_{1}/Q_{0}|$ of $0.87$, $0.92$ and
$1.04$, although large, are consistent with the expected value of
$0.58$ for this quantity. For comparison, a description of $^{109}{\rm
  Ag}$ as a proton-hole coupled to a $^{110}{\rm Cd}$ core was also
performed.  This description is consistent with the one describing
$^{109}{\rm Ag}$ as a proton coupled to a $^{108}{\rm Pd}$ core. In
the former case $|Q_{1}/Q_{0}|=1.23$, probably larger than naively
expected from the EFT.

\begin{figure}
\centering
\includegraphics[width=0.49\textwidth]{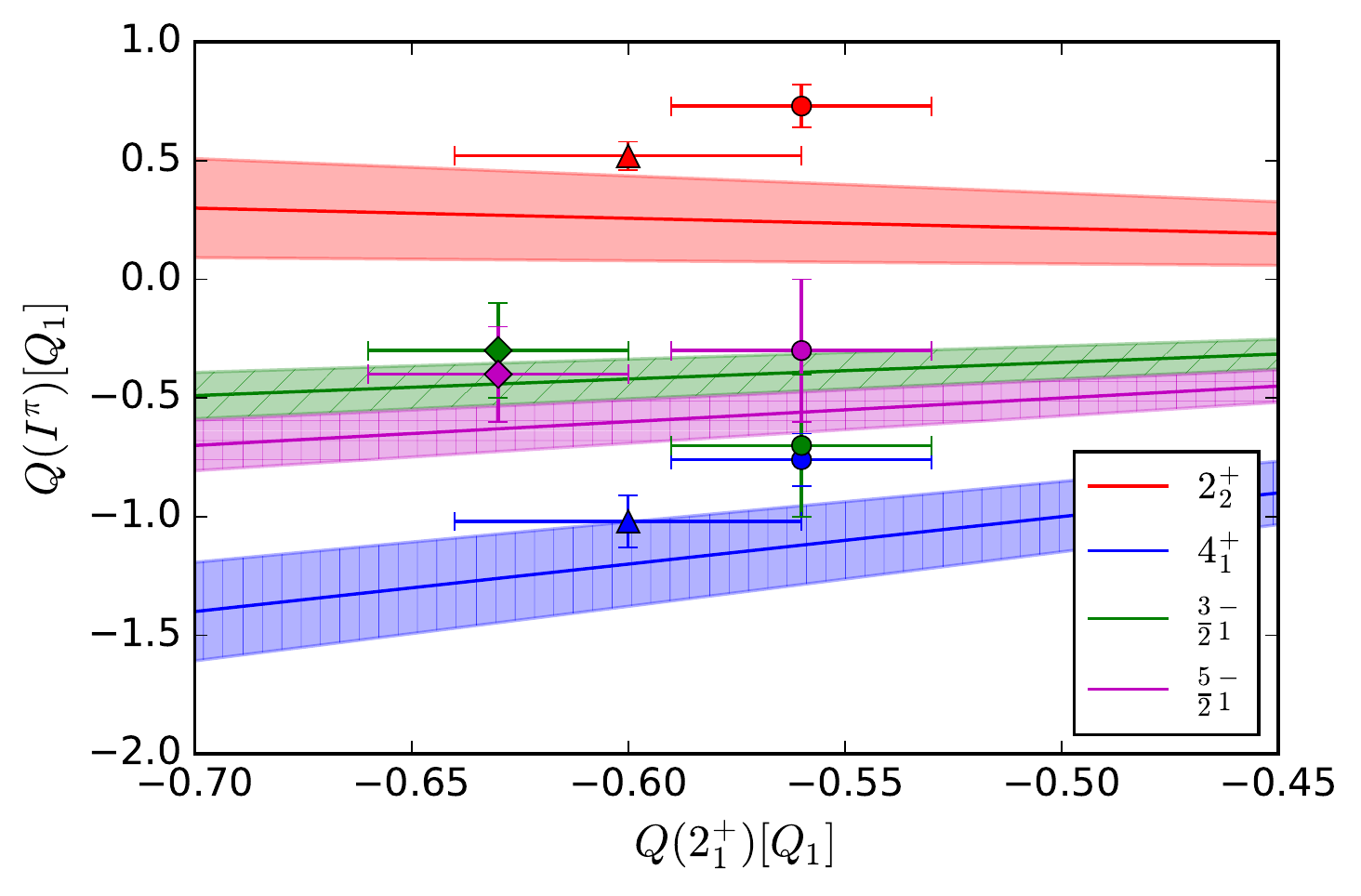}
\includegraphics[width=0.49\textwidth]{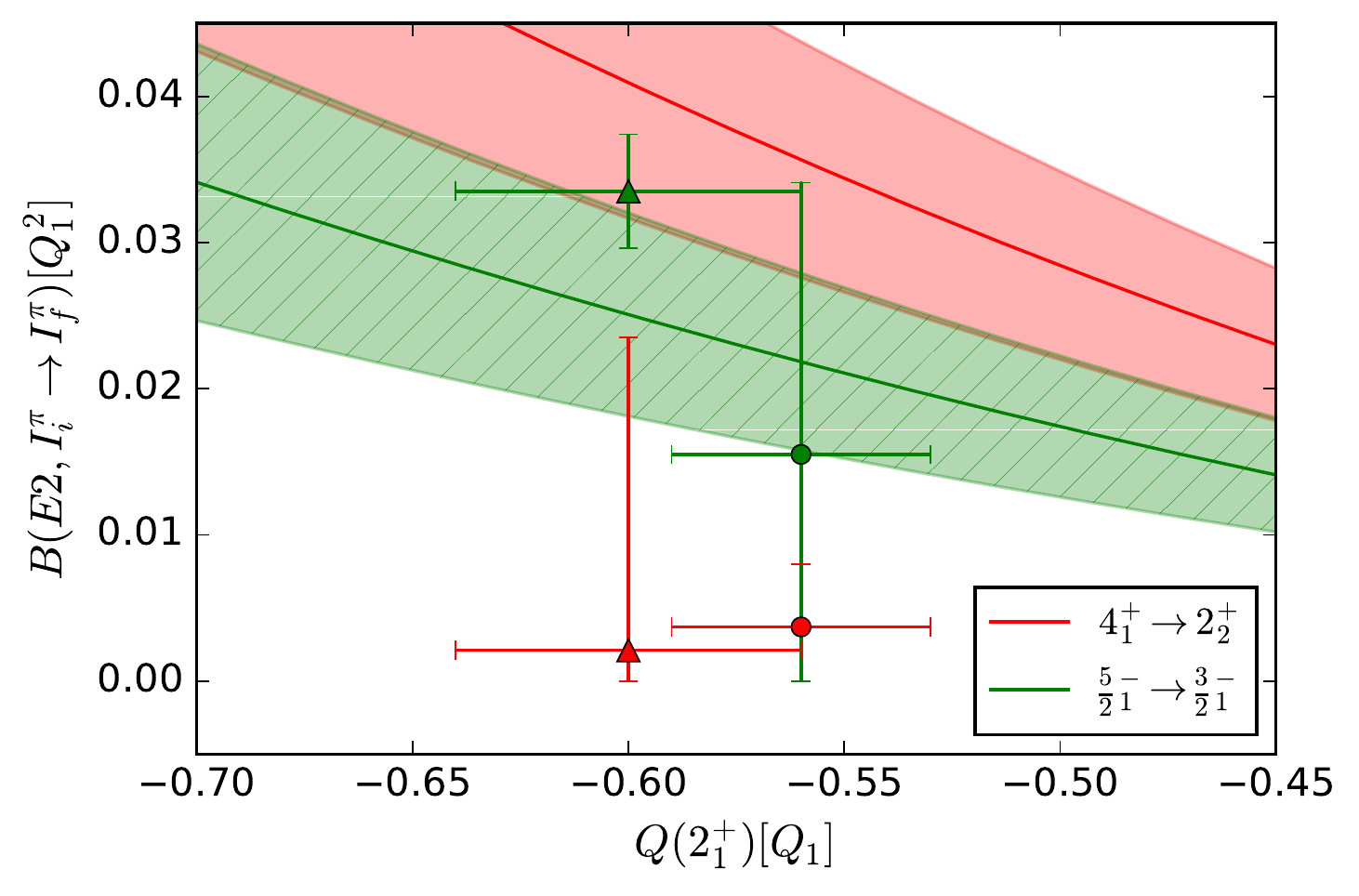}
\caption{(Color online) Static $E2$ moments (top) and $E2$ transitions
  strengths (bottom) as functions of $Q(2_{1}^{+})$. The uncertainty
  quantified from 68\% DOB intervals is shown as error bands. Data for
  the $^{102}{\rm Ru}$/$^{103}{\rm Rh}$, $^{106}{\rm Pd}$/$^{107}{\rm
    Ag}$ and $^{108}{\rm Pd}$/$^{109}{\rm Ag}$ systems are shown as
  diamonds, triangles, and circles, respectively.}
\label{E2(q2)}
\end{figure}

The expressions in Tables~\ref{E2mom} and~\ref{NLOredE2} can be used
to relate different $E2$ observables. As examples, static $E2$ moments
and $E2$ transition strengths are plotted as functions of $Q(2_{1}^{+})$
in Figure~\ref{E2(q2)}.
The theoretical uncertainties associated to these quantities,
quantified using 68\% DOB intervals, are represented by bands. In the
top part of the figure, the static $E2$ moments of the $2_{2}^{+}$,
$4_{1}^{+}$, $(\sfrac{3}{2})_{1}^{-}$ and $(\sfrac{5}{2})_{1}^{-}$
states are shown as red, blue, green and purple lines,
respectively. In the bottom of the figure, the $E2$ transition
strengths for the $4_{1}^{+}\rightarrow 2_{2}^{+}$ and
$(\sfrac{5}{2})_{1}^{-}\rightarrow (\sfrac{3}{2})_{1}^{-}$ transitions
are shown as red and green lines, respectively. Experimental data for
the $^{102}{\rm Ru}$/$^{103}{\rm Rh}$, $^{106}{\rm Pd}$/$^{107}{\rm
  Ag}$ and $^{108}{\rm Pd}$/$^{109}{\rm Ag}$ systems are shown in the
figure as colored diamonds, triangles and circles,
respectively. 
For these systems, the relations plotted in Figure~\ref{E2(q2)} are
fulfilled except for the ones involving the $2_{2}^{+}$ state.

\section{\textit{M}1 observables}\label{M1}
The magnetic dipole ($M1$) operator is a spherical tensor of
rank one. In our EFT, the simplest rank-one operator is
\begin{equation}
\begin{split}
\hat{\mu}_{\mu} = & \mu_{d}\hat{\mathbf{J}}_{\mu} + \mu_{a}\hat{\mathbf{j}}_{\mu} \\
& + \left( \left(d^{\dagger}+\tilde{d}\right) \otimes \left( \mu_{d1}\hat{\mathbf{J}} + \mu_{a1}\hat{\mathbf{j}} \right) \right)^{(1)}_{\mu}.
\end{split}
\label{M1op}
\end{equation}
The first and second terms on the right-hand side of Eq.~(\ref{M1op})
preserve the phonon number, and enter in the LO calculation of static $M1$
moments and phonon-conserving $M1$ transition strengths. The last two
terms enter in the LO calculation of phonon-changing $M1$ transition
strengths. 

Experimental data show that the typical size for the static $M1$ moment
of the even-even $2_{1}^{+}$ state is about one nuclear magneton
$\mu_{N}$. This observation and the fact that in even-even nuclei
\begin{equation}
\langle I || \hat{\mathbf{J}} || I \rangle = \sqrt{I(I+1)(2I+1)},
\end{equation}
allow us to estimate the scale for the LEC $\mu_{d}$ in Eq.~(\ref{M1op}) as
\begin{equation}
\mu_{d} \sim \frac{1}{5}\mu_{N}.
\label{M1scale}
\end{equation}

The Schmidt value for the magnetic moment of a proton in a
$j^{\pi}=\sfrac{1}{2}^{-}$ orbital is $\mu_{p}\approx-0.26\mu_{N}$.
In contrast to $E2$ phenomena, magnetic properties in vibrational
nuclei are not collective, and the contributions of the odd fermion
cannot be neglected. As will be shown in what follows, the static $M1$
moment of the $I=\sfrac{1}{2}$ ground state of the odd-mass nuclei
calculated from the operator~(\ref{M1op}) is $\mu(\sfrac{1}{2})=
\sqrt{\pi/3}\mu_{a}$. Thus, we naively estimate the value of $\mu_{a}$ as
\begin{equation}
\mu_{a}\sim\mu_{p}.
\end{equation}
Static $M1$ moments for the ground state in $^{103}{\rm Rh}$,
$^{107}{\rm Ag}$ and $^{109}{\rm Ag}$ are consistent with this
estimate. It is important to realize that the LEC $\mu_a$ is neither
equal nor simply related to the Schmidt value. In the EFT
considered in this work, we couple a fermion with $j^\pi=\sfrac{1}{2}^-$ (and
not a free proton in a $p$ wave) to a collective state.  We
have no information about any radial wave function of the coupled
fermion, and we have no operators to act on its spin and its orbital
angular momentum separately.  The coupling between the fermion and the
core is strong (as the separation energy $S$ considerably exceeds the
energy scale $\omega$ of core excitations).  The result of the
coupling is again a collective state, and renormalizations replace
``bare'' quantities such as the proton's magnetic moment by effective
couplings.  It is useful to contrast the EFT for vibrations in
odd-mass nuclei with halo EFT~\cite{bertulani2002, higa2008,
  hammer2011, ryberg2014-1} for odd-mass nuclei.  In halo EFT, a
nucleon is very weakly bound to a core, and $S\ll\omega$ holds. The
nucleon's Schmidt value is the leading contribution to the total
magnetic moment, and subleading corrections are of size $S/\omega\ll
1$~\cite{fernando2012, fernando2015}.

Let us now turn to the phonon-changing terms in Eq.~(\ref{M1op}) and
discuss the size of the LECs $\mu_{d1}$ and $\mu_{a1}$. Due to the absence
of strong collective effects in $M1$ observables, the naive expectation is that
transition matrix elements again are of single-particle size, i.e. similar to
$\mu_N$ or $\mu_p$. Higher-order corrections to the leading
phonon-changing and phonon-preserving terms of the $M1$
operator~(\ref{M1op}) enter with increasing powers of boson or fermion
creation and annihilation operators. We expect them to scale as
$\varepsilon$ and omit them in what follows.

The reduced transition probabilities for $M1$ transitions and static $M1$
moments are given by~\cite{rowe2010}
\begin{equation}
B(M1; i\rightarrow f) = \frac{\left|\langle f||\hat{\mu}||i\rangle\right|^{2}}{2I_{i}+1}
\end{equation}
and
\begin{equation}
\mu(I)=\sqrt{\frac{4\pi}{3}}\frac{C_{II10}^{II}}{\sqrt{2I+1}} \langle I || \hat{\mu} || I \rangle,
\label{M1momdef}
\end{equation}
respectively.

\subsection{Static moments and phonon-conserving transition strengths}
The LO static $M1$ moments of even-even and odd-mass nuclei can be
calculated from the reduced matrix elements of the first and second
terms of the $M1$ operator~(\ref{M1op}). These are
\begin{widetext}
\begin{equation}
\langle I'; N; 0 || \hat{\mu} || I; N; 0 \rangle = \left\{\begin{array}{l c}
0 &\mbox{for}~N=0 \\ \\
\mu_{d} \sqrt{I(I+1)}\Pi_I &\mbox{for}~N=1 \\ \\
-2\mu_{d} \sqrt{30}\Pi_{I'I} \left\{ \begin{array}{c c c}
  1 & 2 & 2 \\ 2 & I & I' \end{array} \right\} &\mbox{for}~N=2 \end{array} \right.
\end{equation}
\end{widetext}
and
\begin{widetext}
\begin{equation}
\begin{aligned}
\langle I'; N,J'; \tfrac{1}{2} || \hat{\mu} || I; NJ; \tfrac{1}{2} \rangle &=
\left\{\begin{array}{l c}
0 &\mbox{for}~N=0 \\ \\
-\mu_{d} (-1)^{I+\tfrac{1}{2}} \sqrt{30}\Pi_{I'I} \left\{ \begin{array}{c c c}
1 & 2 & 2 \\ \tfrac{1}{2} & I & I' \end{array} \right\} &\mbox{for}~N=1 \\ \\
2\mu_{d} (-1)^{I+\tfrac{1}{2}} \sqrt{30}\Pi_{I'J'IJ}
\left\{ \begin{array}{c c c} 1 & 2 & 2 \\ 2 & J' & J \end{array} \right\}
\left\{ \begin{array}{c c c} 1 & J' & J \\ \tfrac{1}{2} & I & I' \end{array} \right\}&\mbox{for}~N=2 \end{array} \right. \\ \\
& + \left\{\begin{array}{l c}
\mu_{a} \sqrt{\tfrac{3}{2}} &\mbox{for}~N=0 \\ \\
-\mu_{a} (-1)^{I'+\tfrac{1}{2}} \sqrt{\tfrac{3}{2}}\Pi_{I'I}
\left\{ \begin{array}{c c c} 1 & \tfrac{1}{2} & \tfrac{1}{2} \\ 2 & I & I' \end{array} \right\} &\mbox{for}~N=1 \\ \\
-\mu_{a} (-1)^{I'+\tfrac{1}{2}} \sqrt{\tfrac{3}{2}}\Pi_{I'I}
\left\{ \begin{array}{c c c} 1 & \tfrac{1}{2} & \tfrac{1}{2} \\ J & I & I' \end{array} \right\}
\delta_{J'}^{J} &\mbox{for}~N=2 \end{array} \right.
\label{mamatele}
\end{aligned}
\end{equation}
\end{widetext}
Results are listed in Table~\ref{M1mom}.
\begin{table}
\caption{Static $M1$ moments of states up to the two-phonon level in
  terms of $\mu_{d}$ and $\mu_{a}$.}  \centering
\begin{tabular}{ccc}
\hline\hline
System & $I$ & $\mu(I)$ \\
\hline
even-even & $2$ & $4\sqrt{\frac{\pi}{3}}\mu_{d}$ \\
 & $4$ & $8\sqrt{\frac{\pi}{3}}\mu_{d}$ \\
\hline
odd-mass & $\frac{1}{2}$ & $\sqrt{\frac{\pi}{3}}\mu_{a}$ \\
 & $\frac{3}{2}$ & $\frac{18}{5}\sqrt{\frac{\pi}{3}}\mu_{d} - \frac{3}{5}\sqrt{\frac{\pi}{3}}\mu_{a}$ \\
 & $\frac{5}{2}$ & $4\sqrt{\frac{\pi}{3}}\mu_{d} + \sqrt{\frac{\pi}{3}}\mu_{a}$ \\
 & $\frac{7}{2}$ & $\frac{70}{9}\sqrt{\frac{\pi}{3}}\mu_{d} - \frac{7}{9}\sqrt{\frac{\pi}{3}}\mu_{a}$ \\
 & $\frac{9}{2}$ & $8\sqrt{\frac{\pi}{3}}\mu_{d} + \sqrt{\frac{\pi}{3}}\mu_{a}$ \\
\hline\hline
\end{tabular}
\label{M1mom}
\end{table}
These terms of the $M1$ operator~(\ref{M1op}) couple states with the same number
of phonons, and enter in the LO calculation of the allowed phonon-conserving
$M1$ transition strengths in odd-mass nuclei. The reduced matrix elements
employed to calculate these observables are listed in Table~\ref{LOredM1}.
\begin{table}
\caption{Reduced matrix elements relevant for phonon-conserving $M1$
  transitions in terms of $\mu_{d}$ and $\mu_{a}$.}  \centering
\begin{tabular}{ccc}
\hline\hline
System & $I_{i} \rightarrow I_{f}$ & $\langle f||\hat{\mu}||i\rangle$ \\
\hline
odd-mass & $\frac{5}{2} \rightarrow \frac{3}{2}$ & $-\sqrt{\frac{12}{5}}\mu_{d} + \sqrt{\frac{12}{5}}\mu_{a}$ \\
 & $\frac{9}{2} \rightarrow \frac{7}{2}$ & $-\sqrt{\frac{40}{9}}\mu_{d} + \sqrt{\frac{40}{9}}\mu_{a}$ \\
\hline\hline
\end{tabular}
\label{LOredM1}
\end{table}

Our results for static $M1$ moments in the
$^{102}{\rm Ru}$/$^{103}{\rm Rh}$,
$^{106}{\rm Pd}$/$^{107}{\rm Ag}$ and
$^{108}{\rm Pd}$/$^{109}{\rm Ag}$ systems, with uncertainties quantified
as $\sqrt{4\pi/3(2I+1)}C_{II10}^{II}\mu_{d}\delta$ [in agreement with the definition given in Eq.~(\ref{M1momdef})], where $\delta$ comes from
intervals with a 68\% DOB, are listed in Table~\ref{LOM1}.
\begin{table}
\caption{Static $M1$ moments in the $^{102}{\rm Ru}$/$^{103}{\rm
    Rh}$, $^{106}{\rm Pd}$/$^{107}{\rm Ag}$ and $^{108}{\rm
    Pd}$/$^{109}{\rm Ag}$ systems in units of $\mu_{N}$. Values marked
  with an asterisk were employed to fit the LECs. The uncertainty was
  quantified from 68\% DOB intervals.}  \centering
\begin{tabular}{ccd{2.6}d{2.5}}
\hline\hline
Nucleus & $I^{\pi}_{i}$ & \multicolumn{1}{c}{$\mu_{\rm exp}(I^{\pi}_{i})$} & \multicolumn{1}{c}{$\mu_{\rm EFT}(I^{\pi}_{i})$} \\
\hline
$^{102}{\rm Ru}$ & $2_{1}^{+}$ & 0.85(3)^{*} & 0.85(5) \\
 & $2_{2}^{+}$ & & 0.85(10) \\
 & $4_{1}^{+}$ & & 1.70(8) \\
$^{103}{\rm Rh}$ & $\frac{1}{2}_{1}$ & -0.088^{*} & -0.088 \\
 & $\frac{3}{2}_{1}$ & 0.77(7) & 0.81(5) \\
 & $\frac{5}{2}_{1}$ & 1.08(4) & 0.76(5) \\
 & $\frac{7}{2}_{1}$ & 2.0(6) & 1.7(1) \\
 & $\frac{9}{2}_{1}$ & 2.8(5) & 1.6(1) \\
\hline
$^{106}{\rm Pd}$ & $2_{1}^{+}$ & 0.79(2)^{*} & 0.79(5) \\
 & $2_{2}^{+}$ & 0.71(10) & 0.79(10) \\
 & $4_{1}^{+}$ & 1.8(4) & 1.58(8) \\
$^{107}{\rm Ag}$ & $\frac{1}{2}_{1}$ & -0.11^{*} & -0.11 \\
 & $\frac{3}{2}_{1}$ & 0.98(9) & 0.78(5) \\
 & $\frac{5}{2}_{1}$ & 1.02(9) & 0.68(4) \\
 & $\frac{7}{2}_{1}$ & & 1.6(1) \\
 & $\frac{9}{2}_{1}$ & & 1.5(1) \\
\hline
$^{108}{\rm Pd}$ & $2_{1}^{+}$ & 0.71(2)^{*} & 0.71(4) \\
 & $2_{2}^{+}$ & & 0.71(9) \\
 & $4_{1}^{+}$ & & 1.42(7) \\
$^{109}{\rm Ag}$ & $\frac{1}{2}_{1}$ & -0.13^{*} & -0.13 \\
 & $\frac{3}{2}_{1}$ & 1.10(10) & 0.72(5) \\
 & $\frac{5}{2}_{1}$ & 0.85(8) & 0.58(4) \\
 & $\frac{7}{2}_{1}$ & & 1.5(1) \\
 & $\frac{9}{2}_{1}$ & & 1.3(1) \\
\hline\hline
\end{tabular}
\label{LOM1}
\end{table}
Most experimental values in the table are weighted averages of data
from Refs.~\cite{fahlander1988, svensson1995, stone2014}. The static
$M1$ moment of the $I^{\pi}=\sfrac{1}{2}^{-}$ ground state in
$^{103}{\rm Rh}$ was taken from Ref.~\cite{sogo1955}. For each system,
we adjusted the LECs $\mu_d$ and $\mu_a$ to the static $M1$ moments
of the even-even $2_1^+$ and odd-mass $(\sfrac{1}{2})_{1}^{\pi}$ states,
respectively. Notice that the values for $\mu_{d}$ of $0.21$, $0.19$
and $0.17\mu_{N}$ in the $^{102}{\rm Ru}$/$^{103}{\rm Rh}$,
$^{106}{\rm Pd}$/$^{107}{\rm Ag}$ and
$^{108}{\rm Pd}$/$^{109}{\rm Ag}$ systems, respectively, are all
consistent with the naive estimate $0.2\mu_{N}$. Similarly, the values
for $\mu_{a}$ of $-0.09$, $-0.11$ and $-0.13\mu_{N}$ are all consistent
with the Schmidt value $\mu_{p}=-0.26\mu_{N}$.

Table~\ref{NLOBM1} lists our results for phonon-conserving $M1$
transition strengths in the studied odd-mass nuclei, with uncertainties
quantified as $\mu_{d}^{2}\delta/(2I_{i}+1)$, where $\delta$ comes from
intervals with a 68\% DOB.
\begin{table}
\caption{Reduced transition probabilities for phonon-conserving $M1$
  transitions in Weisskopf units. The uncertainty was quantified from
  68\% DOB intervals.}  \centering
\begin{tabular}{cccc}
\hline\hline
Nucleus & $I^{\pi}_{i} \rightarrow I^{\pi}_{f}$ & $B(M1)_{\rm exp}$ & $B(M1)_{\rm EFT}$ \\
\hline
$^{103}{\rm Rh}$ & $\frac{5}{2}_{1}^{-} \rightarrow \frac{3}{2}_{1}^{-}$ & & $0.034(2)$ \\
 & $\frac{5}{2}_{2}^{-} \rightarrow \frac{3}{2}_{2}^{-}$ & & $0.034(5)$ \\
 & $\frac{9}{2}_{1}^{-} \rightarrow \frac{7}{2}_{1}^{-}$ & & $0.038(3)$ \\
\hline
$^{107}{\rm Ag}$ & $\frac{5}{2}_{1}^{-} \rightarrow \frac{3}{2}_{1}^{-}$ & $0.033(4)$ & $0.036(2)$ \\
 & $\frac{5}{2}_{2}^{-} \rightarrow \frac{3}{2}_{2}^{-}$ & & $0.036(4)$ \\
 & $\frac{9}{2}_{1}^{-} \rightarrow \frac{7}{2}_{1}^{-}$ & & $0.040(2)$ \\
\hline
$^{109}{\rm Ag}$ & $\frac{5}{2}_{1}^{-} \rightarrow \frac{3}{2}_{1}^{-}$ & $0.043(7)$ & $0.036(2)$ \\
 & $\frac{5}{2}_{2}^{-} \rightarrow \frac{3}{2}_{2}^{-}$ & & $0.036(3)$ \\
 & $\frac{9}{2}_{1}^{-} \rightarrow \frac{7}{2}_{1}^{-}$ & & $0.040(2)$ \\
\hline\hline
\end{tabular}
\label{NLOBM1}
\end{table}
The sparse available data on phonon-conserving $M1$ transition strengths
are consistent with the EFT predictions.

\subsection{Phonon-annihilating transition strengths}
The last two terms of the $M1$ operator~(\ref{M1op}) couple states whose
number of phonons differ by one. Their reduced matrix elements allow us
to calculate phonon-annihilating $M1$ transition strengths at LO. In
even-even nuclei, these transitions are higher order effects, as discussed in
Ref.~\cite{coelloperez2015-2}. The reduced matrix elements of these terms
in odd-mass nuclei are
\begin{widetext}
\begin{equation}
\begin{gathered}
\begin{split}
\langle I'; N-1,J'; \tfrac{1}{2} || \hat{\mu} || I; NJ; \tfrac{1}{2} \rangle = & \left\{\begin{array}{l c}
0 &\mbox{for}~N=1 \\ \\
6\mu_{d1} (-1)^{I+\tfrac{1}{2}} \sqrt{5}\Pi_{I'IJ} \left\{ \begin{array}{c c c}
2 & 2 & J \\ 2 & 1 & 1 \end{array} \right\} \left\{ \begin{array}{c c c}
I & I' & 1 \\ 2 & J & \tfrac{1}{2} \end{array} \right\} &\mbox{for}~N=2 \end{array} \right. \\ \\
& + \left\{\begin{array}{l c}
-\mu_{a1} (-1)^{I+j} \sqrt{\frac{9}{2}}\Pi_I \left\{ \begin{array}{c c c} 1 & 1 & 2 \\
\tfrac{1}{2} & I & I' \end{array} \right\} &\mbox{for}~N=1 \\ \\
3 \mu_{a}\Pi_{I'IJ}
\left\{ \begin{array}{c c c} I' & 2 & \tfrac{1}{2} \\  I & J & \tfrac{1}{2} \\
1 & 2 & 1 \end{array} \right\} &\mbox{for}~N=2 \end{array} \right.
\end{split}
\end{gathered}
\label{ma1matele}
\end{equation}
\end{widetext}
The relevant matrix elements for the calculation of these observables
in odd-mass nuclei are listed in Table~\ref{phanredM1}.

\begin{table}
\caption{Reduced matrix elements relevant for phonon-annihilating $M1$ transitions
in terms of $\mu_{d1}$ and $\mu_{a1}$.}
\centering
\begin{tabular}{ccc}
\hline\hline
System & $I_{i} \rightarrow I_{f}$ & $\langle f||\hat{\mu}||i\rangle$ \\
\hline
odd-mass & $\frac{3}{2}_{1} \rightarrow \frac{1}{2}_{1}$ & $-\sqrt{\frac{3}{2}}\mu_{a1}$ \\
 & $\frac{1}{2}_{2} \rightarrow \frac{3}{2}_{1}$ & $\sqrt{\frac{3}{5}}\mu_{a1}$ \\
 & $\frac{3}{2}_{2} \rightarrow \frac{3}{2}_{1}$ & $-\frac{3}{5}\sqrt{42}\mu_{d1} + \frac{1}{5}\sqrt{42}\mu_{a1}$ \\
 & $\frac{3}{2}_{2} \rightarrow \frac{5}{2}_{1}$ & $-\frac{1}{5}\sqrt{42}\mu_{d1} - \frac{1}{10}\sqrt{42}\mu_{a1}$ \\
 & $\frac{5}{2}_{2} \rightarrow \frac{3}{2}_{1}$ & $\frac{1}{5}\sqrt{42}\mu_{d1} + \frac{1}{10}\sqrt{42}\mu_{a1}$ \\
 & $\frac{5}{2}_{2} \rightarrow \frac{5}{2}_{1}$ & $- \frac{14}{5}\sqrt{3}\mu_{d1} - \frac{2}{5}\sqrt{3}\mu_{a1}$ \\
 & $\frac{7}{2}_{1} \rightarrow \frac{5}{2}_{1}$ & $-\sqrt{\frac{27}{5}}\mu_{a1}$ \\
\hline\hline
\end{tabular}
\label{phanredM1}
\end{table}

In Table~\ref{LOBM1} we present our results for phonon-annihilating
$M1$ transition strengths in $^{103}{\rm Rh}$ and $^{109}{\rm Ag}$,
with uncertainties quantified as $\mu_{a1}^{2}\delta/(2I_{i}+1)$,
where $\delta$ comes from 68\% DOB intervals. All available data from
Refs.~\cite{defrenne2009-2, blachot2006} were employed to fit the
LECs.  For $^{103}{\rm Rh}$ and $^{109}{\rm Ag}$ we find values for
$\mu_{d1}$ of $0.0\mu_N$ and $0.08\mu_{N}$, and values for $\mu_{a1}$
of $0.68\mu_N$ and $0.76\mu_{N}$, respectively. The small values for
$\mu_{d1}$, although smaller than naively expected, reflect the fact
that $M1$ transitions in even-even nuclei are higher order
effects. The values for $\mu_{a1}$ are consistent with the naive
estimates. Our results are in agreement with the sparse experimental
data on phonon-annihilating $M1$ transition strengths.

\begin{table}
\caption{Reduced transition probabilities for phonon-annihilating $M1$
  transitions in $^{103}{\rm Rh}$ and $^{109}{\rm
    Ag}$ in Weisskopf units. The uncertainty was quantified from 68\%
  DOB intervals.}  \centering
\begin{tabular}{ccd{2.7}d{2.7}}
\hline\hline
Nucleus & $I^{\pi}_{i}\rightarrow I^{\pi}_{f}$ & \multicolumn{1}{c}{$B(M1)_{\rm exp}$} & \multicolumn{1}{c}{$B(M1)_{\rm EFT}$} \\
\hline
$^{103}{\rm Rh}$ & $\frac{3}{2}^{-}_{1}\rightarrow\frac{1}{2}^{-}_{1}$ & 0.12(1) & 0.10(2) \\
 & $\frac{1}{2}^{-}_{2}\rightarrow\frac{3}{2}^{-}_{1}$ & & 0.08(8) \\
 & $\frac{3}{2}^{-}_{2}\rightarrow\frac{3}{2}^{-}_{1}$ & & 0.10(4) \\
 & $\frac{3}{2}^{-}_{2}\rightarrow\frac{5}{2}^{-}_{1}$ & & 0.03(4) \\
 & $\frac{5}{2}^{-}_{2}\rightarrow\frac{3}{2}^{-}_{1}$ & 0.014(2) & 0.018(28) \\
 & $\frac{5}{2}^{-}_{2}\rightarrow\frac{5}{2}^{-}_{1}$ & 0.020(3) & 0.023(28) \\
 & $\frac{7}{2}^{-}_{1}\rightarrow\frac{5}{2}^{-}_{1}$ & & 0.17(2) \\
\hline
$^{109}{\rm Ag}$ & $\frac{3}{2}^{-}_{1}\rightarrow\frac{1}{2}^{-}_{1}$ & 0.117(15) & 0.122(27) \\
 & $\frac{1}{2}^{-}_{2}\rightarrow\frac{3}{2}^{-}_{1}$ & & 0.10(11) \\
 & $\frac{3}{2}^{-}_{2}\rightarrow\frac{3}{2}^{-}_{1}$ & 0.16(7) & 0.07(5) \\
 & $\frac{3}{2}^{-}_{2}\rightarrow\frac{5}{2}^{-}_{1}$ & & 0.05(5) \\
 & $\frac{5}{2}^{-}_{2}\rightarrow\frac{3}{2}^{-}_{1}$ & 0.036(16) & 0.033(36) \\
 & $\frac{5}{2}^{-}_{2}\rightarrow\frac{5}{2}^{-}_{1}$ & 0.10(4) & 0.07(4) \\
 & $\frac{7}{2}^{-}_{1}\rightarrow\frac{5}{2}^{-}_{1}$ & & 0.22(3) \\
\hline\hline
\end{tabular}
\label{LOBM1}
\end{table}

\section{Discussion of odd-mass cadmium isotopes}
\label{discuss}

The results presented for spectra, $E2$ moments and transitions, and
$M1$ moments and transitions suggest that an EFT approach to odd-mass
nuclei yields a consistent description of low-energy data. Admittedly,
the agreement between theory and data is also due to the relatively
large experimental and theoretical uncertainties. More precise data is
necessary to really probe the theory, and to motivate the computation
of higher-order corrections.

Technically, the EFT we considered falls in the category of
``particle-vibrator'' models.  Very recently, Stuchbery {\it et
  al.}~\cite{stuchbery2016} measured $g$ factors of the odd isotopes
$^{111,113}$Cd and found that the specific particle-vibrator model of
Ref.~\cite{choudhury1954} failed to capture key aspects of the data. A
second attempt to describe these cadmium isotopes was then made within
the particle-rotor (PR) model described in Ref.~\cite{semmes1991}.

What would an EFT approach yield for these isotopes? The
$^{111,113}$Cd nuclei have $I^\pi=\sfrac{1}{2}^+$ ground states, and
some low-lying levels can be viewed as the result of a $j^\pi=
\sfrac{1}{2}^+$ neutron coupled to the collective excitations of
$^{110,112}$Cd. In addition to the $j^\pi=\sfrac{1}{2}^+$ orbital, one
also has to include a very low-lying $j^\pi=\sfrac{5}{2}^+$ orbital in the
description.  Let the fermion creation operators $a^\dagger_\nu$ with
$\nu=-\sfrac{1}{2}, \sfrac{1}{2}$ and $b_\mu^\dagger$ with
$\mu=-\sfrac{5}{2}, -\sfrac{3}{2}, \ldots, \sfrac{5}{2}$ create a fermion
in the $j^\pi=\sfrac{1}{2}^+$ and $j^\pi=\sfrac{5}{2}^+$ orbital,
respectively.  The LO Hamiltonian that governs the interactions between
the fermion degrees of freedom and the quadrupole bosons is
\begin{eqnarray}
\label{hamcd}
H_{\rm abd} &=& -S(\hat{n}_a+\hat{n}_b) \nonumber\\
&&+\omega_1 \hat{N} + \omega_b\hat{n}_b \nonumber\\
&&+g_{da}\hat{\mathbf{J}}\cdot\hat{\mathbf{j}}_{a} + 
g_{db}\hat{\mathbf{J}}\cdot\hat{\mathbf{j}}_{b} \nonumber\\
&&+\omega_{2a} \hat{N}\hat{n}_a + \omega_{2b} \hat{N}\hat{n}_b .
\end{eqnarray}
Here, we used the operators
\begin{eqnarray}
\hat{n}_a &\equiv& a^\dagger\cdot\tilde{a} , \\
\hat{n}_b &\equiv& b^\dagger\cdot\tilde{b} , \\
\hat{\mathbf{j}}_a &\equiv& {1\over\sqrt{2}}\left(a^\dagger\otimes\tilde{a}\right)^{(1)}  , \\
\hat{\mathbf{j}}_b &\equiv& {\sqrt{70}\over 2}\left(b^\dagger\otimes\tilde{b}\right)^{(1)}  .
\end{eqnarray}
In the Hamiltonian~(\ref{hamcd}) we omitted terms that are quartic in
the boson operators.  As before, $S$ denotes the separation energy and
is the largest energy scale in the Hamiltonian. The difference between
the separation energies of the $a$ and $b$ fermions is denoted as
$\omega_b\approx0.3$~MeV, and is similar in size as
$\omega_1$. Interactions between the fermion orbitals are smaller
corrections and omitted. The Hamiltonian~(\ref{hamcd}) simply
describes two fermion orbitals that interact with the quadrupole
bosons but do not interact with each other. Its eigenstates are simple
product states.

Within this EFT, the phonon-conserving part of the $M1$ operator has
the leading terms
\begin{equation}
\hat{\mu} = \mu_d \hat{\mathbf{J}} + \mu_a\hat{\mathbf{j}}_a + \mu_b\hat{\mathbf{j}}_b  .
\end{equation}
\citeauthor{stuchbery2016} found the static $M1$ moments of the
ground state $|(\sfrac{1}{2})_{1}^{+}\rangle=a^\dagger|0\rangle$ and the
excited states
\begin{equation}
| \left(\tfrac{5}{2}\right)^{+}_{1} \rangle = b^{\dagger}|0\rangle
\quad {\rm and} \quad
| I_{f}^{+}\rangle = \left(d^\dagger \otimes f^\dagger\right)^{(I)}
|0\rangle,
\label{excited}
\end{equation}
with $f=a,b$ and $I=\sfrac{3}{2}, \sfrac{5}{2}$, of particular interest.
For these states we have
\begin{equation}
\begin{aligned}
\langle \left(\tfrac{1}{2}\right)^{+}_{1} || \hat{\mu} ||
\left(\tfrac{1}{2}\right)^{+}_{1}\rangle = &
\mu_{a} \sqrt{\frac{3}{2}} \\
\langle \left(\tfrac{5}{2}\right)^{+}_{1} || \hat{\mu} ||
\left(\tfrac{5}{2}\right)^{+}_{1}\rangle = &
\mu_{b} \sqrt{\frac{105}{2}} \\
\langle I^{+}_{f} || \hat{\mu} || I^{+}_{f}\rangle = &
\mu_{d}\Pi_{i}\frac{I(I+1)-F(j_{f})}{2\sqrt{I(I+1)}} \\
& + \mu_{f}\Pi_{I}\frac{I(I+1)+F(j_{f})}{2\sqrt{I(I+1)}}.
\end{aligned}
\label{matexcited}
\end{equation}
Here $F(j_{f})\equiv j_{f}(j_{f}+1)-6$. The static $M1$ moments of
$I^{\pi}=\sfrac{1}{2}^{+}, \sfrac{3}{2}^{+}, \sfrac{5}{2}^{+}$ states in
odd-mass cadmium isotopes that result from the coupling of the $a$ neutron
to the even-even core are given by the expressions listed in Table~\ref{M1mom}.
The static $M1$ moments of state resulting from the coupling of the $b$ neutron
to the core are
\begin{equation}
\begin{aligned}
\mu\left(\left(\tfrac{5}{2}\right)_{1}^{+}\right) = & 5\sqrt{\frac{\pi}{3}}\mu_{b},\\
\mu\left(\left(\tfrac{3}{2}\right)_{b}^{+}\right) = & \frac{2}{5}\sqrt{\frac{\pi}{3}}\mu_{d}
+ \frac{13}{5}\sqrt{\frac{\pi}{3}}\mu_{b}\\
\mu\left(\left(\tfrac{5}{2}\right)_{b}^{+}\right) = & \frac{12}{7}\sqrt{\frac{\pi}{3}}\mu_{d}
+ \frac{23}{7}\sqrt{\frac{\pi}{3}}\mu_{b}
\end{aligned}
\end{equation}

One can adjust the LECs $\mu_{d}$, $\mu_{a}$ and $\mu_{b}$ to the
static $M1$ moments of the even-even $2^{+}_{1}$ and odd-mass
$(\sfrac{1}{2})^{+}_{1}$ and $(\sfrac{5}{2})^{+}_{1}$ states,
respectively, and predict the static $M1$ moments of the rest of the
excited states.  Our results for the static $M1$ moments in the
$^{110}{\rm Cd}$/$^{111}{\rm Cd}$ and $^{112}{\rm Cd}$/$^{113}{\rm
  Cd}$ systems are listed in Table~\ref{Cd} together with those of
Ref.~\cite{stuchbery2016} calculated within the PR model of
Ref.~\cite{semmes1991}.
\begin{table}
\caption{Static $M1$ moments in the $^{110}{\rm Cd}$/$^{111}{\rm Cd}$
  and $^{112}{\rm Cd}$/$^{113}{\rm Cd}$ systems in units of $\mu_{N}$.
  The static $M1$ moments labeled as $\mu_{\rm PR}$ were taken from
  Ref.~\cite{stuchbery2016} and calculated within the PR model of
  Ref.~\cite{semmes1991}. Values marked with an asterisk were employed
  to fit the LECs of the EFT. The uncertainty was quantified from 68\%
  DOB intervals.}  \centering
\begin{tabular}{ccd{2.6}d{2.6}d{2.6}}
\hline\hline
Nucleus & $I^{\pi}_{i}$ & \multicolumn{1}{c}{$\mu_{\rm PR}(I^{\pi}_{i})$} & \multicolumn{1}{c}{$\mu_{\rm exp}(I^{\pi}_{i})$} & \multicolumn{1}{c}{$\mu_{\rm EFT}(I^{\pi}_{i})$} \\
\hline
$^{110}{\rm Cd}$ & $2_{1}^{+}$ &  & 0.52(4)^{*} & 0.52(14) \\
 & $2_{2}^{+}$ & & & 0.52(28) \\
 & $4_{1}^{+}$ & & & 1.0(2) \\
$^{111}{\rm Cd}$ & $\frac{1}{2}_{1}$ & -0.62 & -0.59^{*} & -0.59 \\
 & $\frac{3}{2}_{1}$ & 0.9 & 0.9(6) & 0.8(1) \\
 & $\frac{5}{2}_{2}$ & 0.8 & 0.5(1) & -0.07(14) \\
\hline
$^{112}{\rm Cd}$ & $2_{1}^{+}$ & & 0.64(16)^{*} & 0.64(15) \\
 & $2_{2}^{+}$ & & & 0.64(30) \\
 & $4_{1}^{+}$ & & & 1.3(2) \\
$^{113}{\rm Cd}$ & $\frac{1}{2}_{1}$ & -0.56 & -0.62^{*} & -0.62 \\
 & $\frac{5}{2}_{1}$ & & -0.77^{*} & -0.77 \\
 & $\frac{3}{2}_{1}$ & 0.8 & -0.6(10) & 0.9(2)\footnotemark[1]  \\
 & & & & -0.3(1)\footnotemark[2] \\
 & $\frac{5}{2}_{2}$ & 0.65 & 0.35(10) & 0.02(14) \\
 & $\frac{3}{2}_{2}$ & 1.2 & 2.1(6) & 0.9(2) \\
\hline\hline
\end{tabular}
\footnotetext[1]{Value obtained assuming the state results from the coupling
of a $j^{\pi}=\sfrac{1}{2}^{+}$ proton to the core.}
\footnotetext[2]{Value obtained assuming the state results from the coupling
of a $j^{\pi}=\sfrac{5}{2}^{+}$ proton to the core.}

\label{Cd}
\end{table}
Theoretical uncertainties were quantified as
$\sqrt{4\pi/3(2I+1)}C_{II10}^{II}\mu_{a}\delta$, where $\delta$ comes
from intervals with a 68\% DOB.  Experimental data for the even-even
nuclei were taken from Refs.~\cite{gurdal2012, defrenne1996}, leading
to values for $\mu_{d}$ of $0.13$ and $0.16\mu_{N}$ in agreement with
the naive expectation for the size of this LEC. Experimental data for
states in the odd-mass nuclei were taken from
Refs.~\cite{stuchbery2016, blachot2009, blachot2010}. Static $M1$
moments were calculated from the $g$ factors of
Ref.~\cite{stuchbery2016} as
\begin{equation}
\mu(I^{\pi}) = g I.
\end{equation}

The values for $\mu_{a}$ of $-0.58$ and $-0.61\mu_{N}$ are small, but
still consistent with the Schmidt value for a neutron in a $j^{\pi}=
\sfrac{1}{2}^{+}$ orbital given by $\mu_{n}\approx-1.91\mu_{N}$.  The
static $M1$ moment of the $(\sfrac{5}{2})_{1}^{+}$ state in
$^{113}{\rm Cd}$ was assumed to be equal to that of the
$(\sfrac{5}{2})_{1}^{+}$ state in $^{111}{\rm Cd}$~\cite{blachot2009}.
Thus, for both cadmium systems $\mu_{b}\approx-0.15\mu_{N}$.  The
static $M1$ moments of the ground states in both odd-mass cadmium
isotopes are well reproduced by the EFT and the PR model, although in
the former case this is attributable to the fact that the static $M1$
moment of the ground state is employed to fit one of the LECs. For
$^{111}{\rm Cd}$, the static $M1$ moment of the
$(\sfrac{3}{2})_{1}^{+}$ state is described by both the EFT and the PR
model. This is not the case for the static $M1$ moment of the
$(\sfrac{5}{2})_{2}^{+}$ state, which is underpredicted by the EFT and
overpredicted by the PR model. For $^{113}{\rm Cd}$, the static $M1$
moment of the $(\sfrac{3}{2})_{1}^{+}$ state is overpredicted by both
the PR model and the EFT unless we assume that this state results from
the coupling of the $b$ neutron to the one-phonon state of the
even-even core. The static $M1$ moment of the $(\sfrac{5}{2})_{2}^{+}$
is underpredicted by the EFT and overpredicted by the PR model, while
the static $M1$ moment of the $(\sfrac{3}{2})_{2}^{+}$ state is
underpredicted by both the EFT and the PR model.  Thus the EFT and the
PR model both yield a fair description of the data.

\section{Summary}\label{summary}
We have developed an EFT for the simultaneous description of spherical
even-even/odd-mass systems in terms of a fermion $j=\sfrac{1}{2}$
degree of freedom coupled to the quadrupole degrees of freedom of the
even-even core. Taking the breakdown scale around the three-phonon
level in the even-even core we systematically expand energies and
electromagnetic observables of states up to the two-phonon level in
terms of the ratio between the corresponding energy and the breakdown
scale.  In the studied odd-mass isotopes of rhodium and silver,
predictions for energy spectra and electromagnetic moments and
transitions strengths are consistent with experimental data within the
theoretical uncertainties quantified via Bayesian methods.  The static
$E2$ moments of excited states and phonon-conserving $E2$ transition
strengths in the even-even and odd-mass nuclei follow the LO relations
predicted by the EFT. While most of the data is consistently described
for LECs of natural size, the strengths of phonon-conserving $M1$
transitions seems to be underpredicted by a factor of about two within
the EFT. More experimental data on these transitions and/or data with
an increased precision would be valuable to further test the EFT
developed in this work.

\begin{acknowledgments}
We thank N. J. Stone and L. Platter for useful discussions. This
material is based upon work supported by the Deutsche
Forschungsgesellschaft under  Grant  SFB  124, and by the U.S. Department of
Energy, Office of Science, Office of Nuclear Physics under Award
Number DEFG02-96ER40963 (University of Tennessee), and under Contract
No.  DE-AC05-00OR22725 (Oak Ridge National Laboratory).
\end{acknowledgments}

\end{document}